\documentclass[twocolumn,prc,aps,showkeys,superscriptaddress,nofootinbib,floatfix,letterpaper]{revtex4}

\usepackage{amssymb}
\usepackage{amsmath}
\usepackage{graphicx} 
\usepackage{epsfig}   
\usepackage{dcolumn}  
\usepackage{rotating} 
\usepackage{subfigure} 
\usepackage{comment}
\usepackage{hyperref}
\usepackage{breakurl}

\usepackage{float}    
\floatstyle{boxed}

\begin{document}

\title{The Aerogel \v{C}erenkov Detector for the SHMS magnetic spectrometer 
in Hall C at Jefferson Lab}

\newcommand*{\CUA}{The Catholic University of America, Washington, DC 20064, USA}

\newcommand*{\ANSL}{A.~I.~Alikhanyan National Science Laboratory, 
Yerevan 0036, Armenia}

\newcommand*{\JLAB}{Thomas Jefferson National 
Accelerator Facility, Newport News, Virginia 23606, USA}

\newcommand*{\MSS}{Mississippi State University, 
Mississippi State, Mississippi 39762, USA }

\newcommand*{\USC}{University of South Carolina, Columbia, 
South Carolina 29208, USA }

\newcommand*{\UIUC}{University of Illinois, Urbana-Champaign, 
Illinois, USA }

\newcommand*{\FIU}{Florida International University, University Park, 
Florida 33199, USA }

\author{T.~Horn} 
\affiliation{\CUA}
\affiliation{\JLAB}

\author{H.~Mkrtchyan}
\affiliation{\ANSL}

\author{S.~Ali} 
\affiliation{\CUA}

\author{A.~Asaturyan}
\affiliation{\ANSL}

\author{M.~Carmignotto} 
\affiliation{\CUA}

\author{A.~Dittmann} 
\affiliation{\UIUC}

\author{D.~Dutta}
\affiliation{\MSS}

\author{R.~Ent}
\affiliation{\JLAB}

\author{N.~Hlavin} 
\affiliation{\CUA}

\author{Y.~Illieva} 
\affiliation{\USC}

\author{A.~Mkrtchyan}
\affiliation{\CUA}

\author{P. Nadel-Turonski}
\affiliation{\JLAB}

\author{I.~Pegg}
\affiliation{\CUA}

\author{A.~Ramos} 
\affiliation{\FIU}

\author{J.~Reinhold} 
\affiliation{\FIU}

\author{I.~Sapkota} 
\affiliation{\CUA}

\author{V.~Tadevosyan} 
\affiliation{\ANSL}

\author{S.~Zhamkochyan}
\affiliation{\ANSL}

\author{S.~A.~Wood}
\affiliation{\JLAB}

\newpage
\date{\today}


\begin{abstract}

Hadronic reactions producing strange quarks such as the exclusive 
$p(e, e^\prime K^+)\Lambda$ and $p(e, e^\prime K^+)\Sigma^0$ reactions, or the 
semi-inclusive $p(e,e^\prime K^+)X$ reaction, play an important role in studies 
of hadron structure and the dynamics that bind  the most basic elements of 
nuclear physics.
The small-angle capability of the new Super High Momentum Spectrometer (SHMS) 
in Hall C, coupled with its high momentum reach - up to the anticipated 11-GeV 
beam energy in Hall~C - and coincidence capability with the well-understood 
High Momentum Spectrometer (HMS), will allow for probes of such hadron 
structure involving strangeness down to the smallest distance scales to date.
To cleanly select the kaons, a threshold aerogel Cerenkov detector has been 
constructed for the SHMS. The detector consists of an aerogel tray followed by 
a diffusion box. Four trays for aerogel of nominal refractive indices
 of $n$=1.030, 1.020, 1.015 and  1.011 were constructed.
The tray combination will allow for identification of kaons from 1 GeV/$c$ up 
to 7.2 GeV/$c$, reaching $\sim 10^{-2}$ proton and $10^{-3}$  pion rejection, 
with kaon  detection efficiency better than 95\%.
The diffusion box of the detector is equipped with 14 five-inch diameter photomultiplier tubes.
Its interior walls are covered with Gore diffusive reflector, which is 
superior to the commonly used Millipore paper and improved the detector 
performance by 35\%. The inner surface of the two aerogel trays with higher 
refractive index is covered with Millipore paper, however, those  
two trays with lower aerogel refractive index are again covered with 
Gore diffusive reflector for higher performance.
The measured mean number of photoelectrons in saturation is $\sim$12 for  
$n$=1.030, $\sim$8 for $n$=1.020, $\sim$10 for $n$=1.015, and $\sim$5.5 for $n$=1.011.
The design details, the results of component characterization, and initial 
performance tests and optimization of the detector are presented.

\end{abstract}

\maketitle

\section{Introduction}
\label{intro}

The 21st century holds great promise for reaching a new era for unlocking the 
mysteries of the inner quark-gluon structure of the atomic nucleus and the 
nucleons inside it as governed by the theory of strong interactions (QCD). 
The experimental program at the upgraded 12 GeV Jefferson Laboratory \cite{NSAC15,mckeown15,dudek12}, currently
 nearing completion, will play an important role in this quest, and 
revolutionize the current understanding of the dynamics of the fundamental 
quarks and gluons that create the wide and varied structure of hadrons and 
nuclei. 

\begin{figure}[H]
\centering
\subfigure[\label{fig-shms-aero} Schematic drawing of the SHMS aerogel detector]
{\includegraphics[width=1.6in]{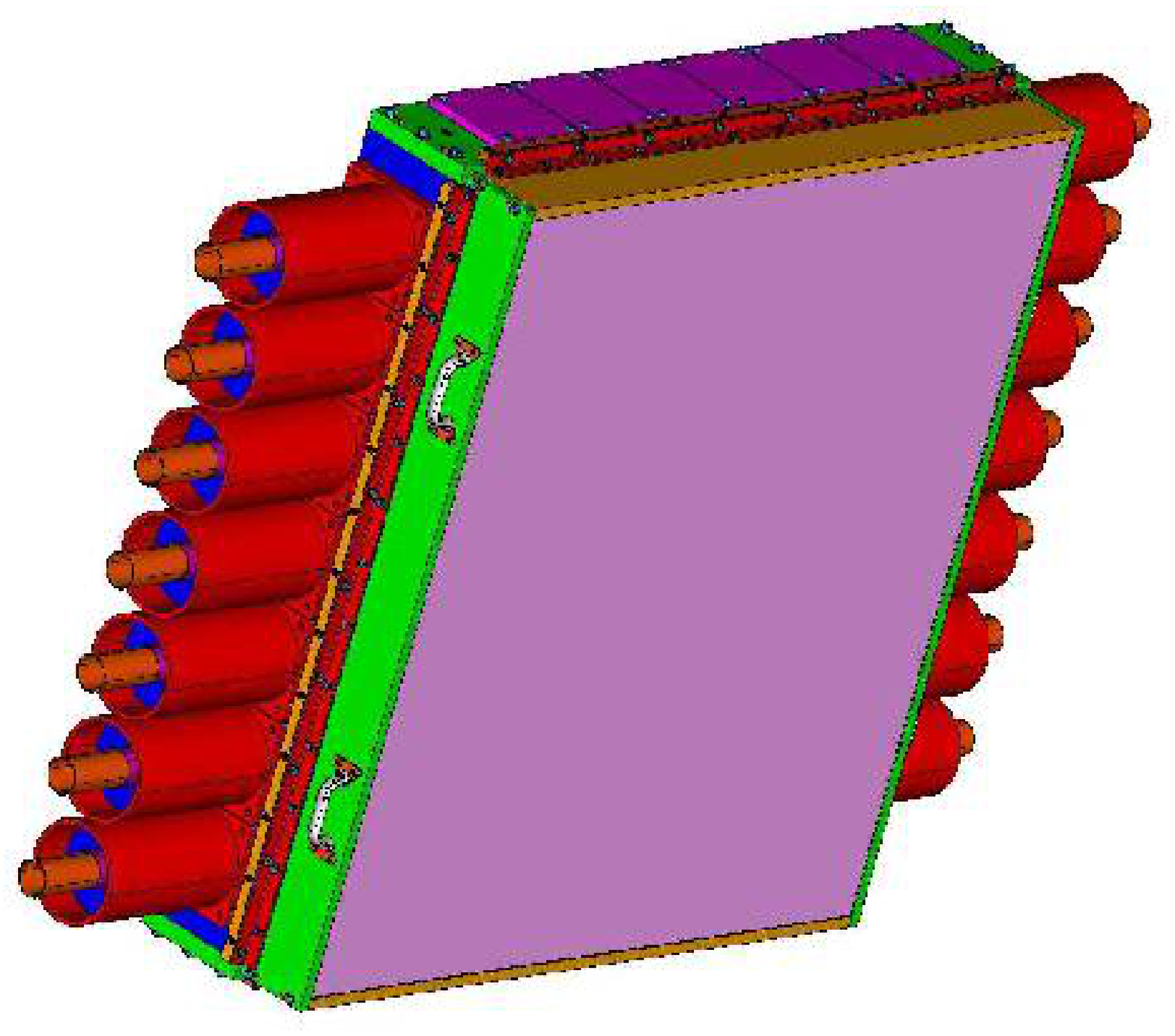} }
\subfigure[\label{shms-hut} Schematic side-view of the SHMS detector package]
{\includegraphics[width=1.6in]{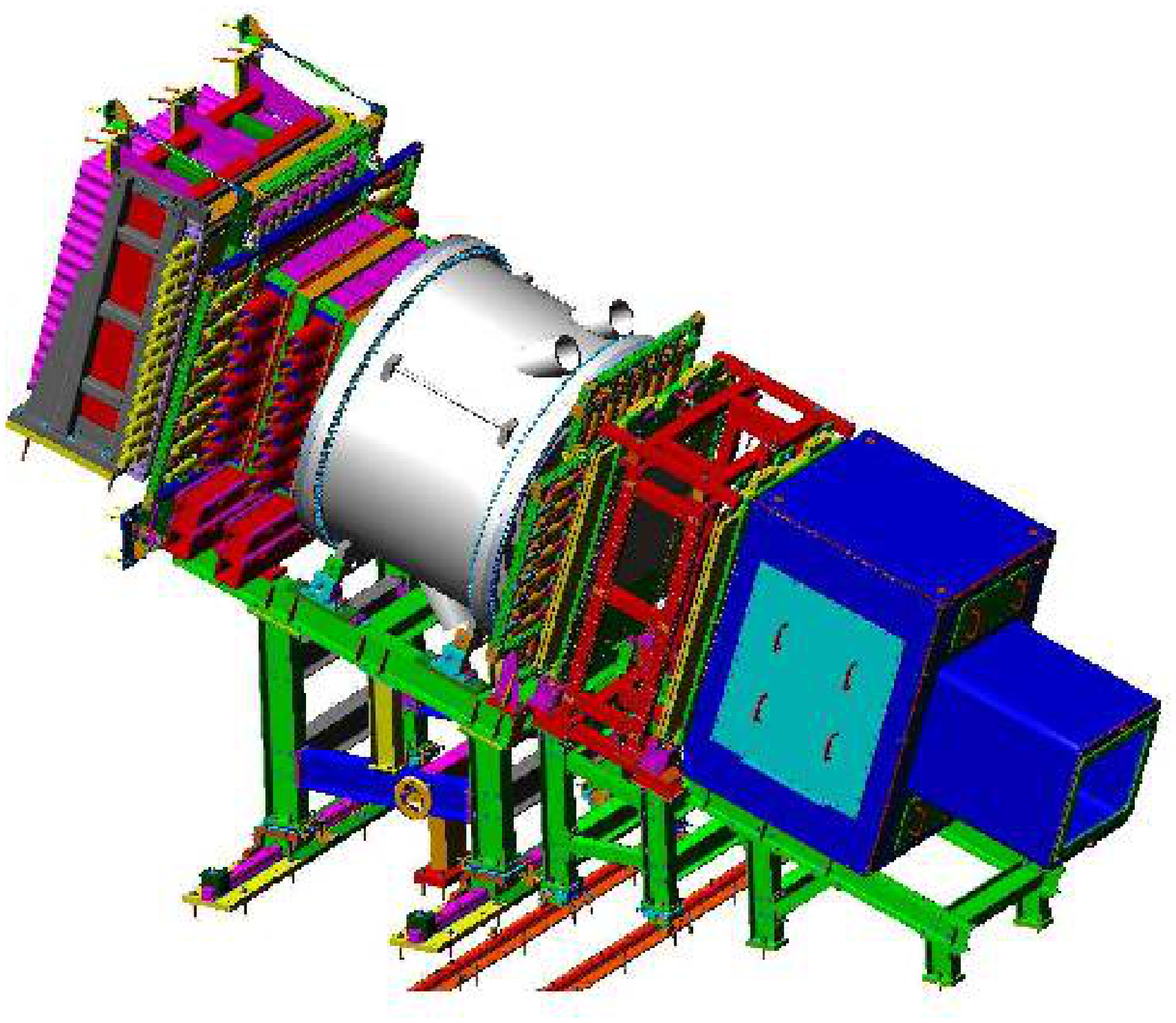}}
\caption{\label{fig-aero-and-shms-hut} (Color online) Schematic drawings of 
the SHMS aerogel detector (left, a) and its placement in the SHMS detector hut 
(right, b). The detector stack of the SHMS consists, right to left, of a gas 
\v{C}erenkov, a pair of drift chambers, a pair of $X-Y$ segmented scintillator,
 a gas \v{C}erenkov (gray cylinder), two possible slots for the aerogel 
detector, each of 30 cm depth, a second $X-Y$ segmented scintillator/quartz bar
hodoscope, and an electromagnetic calorimeter. }
\end{figure}

Hall C with its heavily-shielded detector setup in a highly-focusing magnetic 
spectrometer with large momentum reach will be the optimal Hall for certain 
classes of deep exclusive and semi-inclusive measurements, and in particular 
those requiring high quality Rosenbluth (or longitudinal-transverse - L/T) 
cross section separations. 
The Hall~C base experimental equipment consists of two magnetic spectrometers: 
the High Momentum Spectrometer (HMS) \cite{block08} and the Super High Momentum Spectrometer 
(SHMS) \cite{brindza09}. Depending on the specific requirements of the experiments, they can 
detect either negatively or positively charged particles by choosing the 
magnetic field and the trigger configuration. 
The HMS is designed to detect secondary products of reactions in the momentum 
range from 0.5 to 7.3 GeV/$c$, while the SHMS momentum range extends up to 
about 12 GeV/$c$. 
Both spectrometers are equipped with a pair of drift chambers, pairs of  
timing scintillator hodoscopes for trigger formation, and various detectors for
particle identification purposes, including the aerogel \v{C}erenkov detector.
A schematic of the SHMS detector package is shown in Fig.~\ref{shms-hut}.

Particle identification in the SHMS relies on time-of-flight measurements at 
lower momentum, an electromagnetic pre-shower and full shower calorimeter and 
\v{C}erenkov detectors. 
A \v{C}erenkov detector as filled with noble gases, and the calorimeter will 
be used for $e/\pi$ separation, while a separate \v{C}erenkov detector filled 
with heavier gases such as C$_4$F$_8$O will provide $\pi/K$ identification 
above 3.4 GeV/$c$. 
To complete the necessary PID capability of the SHMS necessary to 
successfully carry out the strangeness physics part of the Hall C program,
the SHMS aerogel \v{C}erenkov detector descriped in this paper was built  
to distinguish kaons from protons with momenta above 2.6 GeV/$c$.

Meson production experiments such as the ``Scaling Study of the L-T Separated 
Pion Electroproduction Cross Section at 11 GeV'' (E12-07-105)~\cite{E12-07-105}
 and ``Studies of the L-T Separated Kaon Electroproduction Cross Section from 
5-11 GeV'' (E12-09-011)~\cite{E12-09-011}, will study the potential for 3D 
hadron imaging studies and may also allow for form factor extractions in a 
kinematic regime where the signatures of QCD are quantitatively revealed~\cite{horn16,favart16}. 
These experiments will provide high-quality cross section data, requiring 
detection of scattered mesons with momenta approaching 10 GeV/$c$, for which 
the SHMS in Hall C is exceptionally well suited.
However, additional instrumentation for $K^+$/$p$ particle identification is 
needed for the kaon measurement, which forms an essential part of this program.
 The ``$P_T$ in SIDIS'' experiment (E12-09-017)~\cite{E12-09-017} will map the 
transverse momentum dependence of semi-inclusive electroproduction of charged 
mesons from both proton and deuteron targets. In this experiment it will be 
beneficial to also separate kaons from protons. 
The ``Measurement of $R$=$\sigma_L$/$\sigma_T$ in SIDIS'' 
(E12-06-104)~\cite{E12-06-104} will study the inclusive-exclusive connection 
in meson electroproduction, where duality has recently been shown to be valid 
and will be essential in understanding flavor decomposition in semi-inclusive 
deep inelastic scattering at 12 GeV. 
With $K^+$/$p$ separation this measurement can access the additional flavors 
of strange systems.
The ``Pion Transparency'' experiment (E12-06-107)~\cite{E12-06-107} will map the
transparency of the nuclear medium to pions and measure the approach to the
QCD-calculable, or point-like, regime using the $Q^2$ dependence of the ratio of
cross sections from nucleon and nuclear targets. Kaon transparency measurements
may contribute additional  information to our understanding of this approach.
The general requirement for these experiments is a high detection efficiency 
for kaons in the SHMS and the capability to separate protons from kaons. 
The experiments plan to run with an electron beam with energy of up to 11 GeV
and beam currents between a few and 80 $\mu$A hitting liquid hydrogen or 
deuterium target cells with lengths about 10 cm, or selected solid targets, 
yielding particle rates in the SHMS detector stack of up to 2 MHz.
To take full advantage of the existing SHMS standard detector configuration 
without compromising its performance, the additional aerogel \v{C}erenkov
detector system was designed with the following specifications:
\begin{itemize}
  \item[-]{to have a sensitive area of at least 90$\times$60 cm$^2$ 
 with a thickness of radiator up to 9~cm.}
  \item[-]{to fit into a small slot of about 30 cm between the heavy gas 
\v{C}erenkov detector and the electromagnetic calorimeter;}
  \item[-]{to minimize the material in the particle path to keep the amount of 
multiple scattering and $\delta$-electrons small;}
  \item[-]{to allow easy exchange of the aerogel material to match the required 
kaon momentum range;}
  \item[-]{to provide acceptable time resolution and a high-rate capability.}
\end{itemize}
Of the approved experiments, three (E12-06-104, E12-09-017, and E12-07-105)
only require kaon particle identification (PID) in the 2.7 - 5 GeV/$c$ 
momentum range.
Higher momenta are relevant for the kaon factorization experiment (E12-09-011),
up to 7.2 GeV/c, and the pion transparency measurement, where kaon 
identification up to this momentum would allow extraction of kaon 
transparencies up to $Q^2 \sim$ 7 (GeV/$c$)$^2$.
To take into account this range of experimental requirements, covering from 
about 2 to above 7~GeV/$c$ momentum, the refractive indices characteristic of 
aerogel material are well suited for this purpose. 

To maintain good $K^+/p$ separation over this full kinematic range, we chose a 
range of aerogel materials, with nominal refractive indices of $n$=1.030 
(SP-30), $n$=1.020 (SP-20), $n$=1.015 (SP-15), and $n$=1.011 (SP-11). 
The threshold momenta (in GeV/$c$) for muons, pions, kaons and protons in these
 types of aerogel are listed in Table~\ref{tab:aerogel-thresholds}.
Up to about 4 GeV/$c$, $K^+/p$ separation is achieved with aerogel with a 
refractive index of $n$=1.030. Beyond, the lower three refractive indices 
provide excellent $K^+/p$ separation in the momentum range between 4 and 
$\sim$7 GeV/$c$.

\begin{table}[H]
\begin{center}
\caption{ Threshold momenta P$_{Th}$ for \v{C}erenkov radiation for charged muons, pions, kaons, and protons in 
aerogel of four refractive indices ranging from $n$=1.011 to 1.030.}
\label{tab:aerogel-thresholds} \smallskip
\small
\begin{tabular}{||l|c|c|c|c|} \hline
Particle & P$_{Th}$   & P$_{Th}$    & P$_{Th}$    & P$_{Th}$    \\
         & $n$=1.030 &  $n$=1.020  & $n$=1.015  &  $n$=1.011 \\
\hline 
 $\mu$   &  0.428    &  0.526      & 0.608      &  0.711     \\
 $\pi$   &  0.565    &  0.692      & 0.803      & 0.935      \\
 $K$     &  2.000    &  2.453      & 2.840      &  3.315     \\
 $p$     &  3.802    &  4.667      & 5.379      &  6.307     \\
\hline
\hline\end{tabular}
\end{center}
\end{table}

This paper is organized as follows. Details regarding the design of the 
detector, such as the diffusion box, tray construction, and a short description 
of the detector components are given in section~\ref{design}.   
Section~\ref{aero_study} presents the results of the aerogel optical properties
based on a randomly selected sample of 5-15\% of the aerogel tiles of each 
refractive index used in the detector. 
Variations in refractive index, dimensions, and surface shape of the 
tiles are also briefly described. 
The construction of a single counter (prototype) and tests performed 
with it are described in section~\ref{aero_study}.
The description of the methods used to study the photomultiplier tubes, 
the results of these studies and the principles guiding selection of the PMTs
for the detector are given in section~\ref{pmt_study}.
The results of initial performance tests of the aerogel \v{C}erenkov detector
and a comparison to simulations are presented in section~\ref{perform}.  
Finally, the results are summarized in section~\ref{summary}.

\section{Design overview}
\label{design}

\begin{figure}[!htbp]
\centering
\subfigure[\label{aero-diff-box}  Assembled diffusion box]
{\includegraphics[width=1.6in]{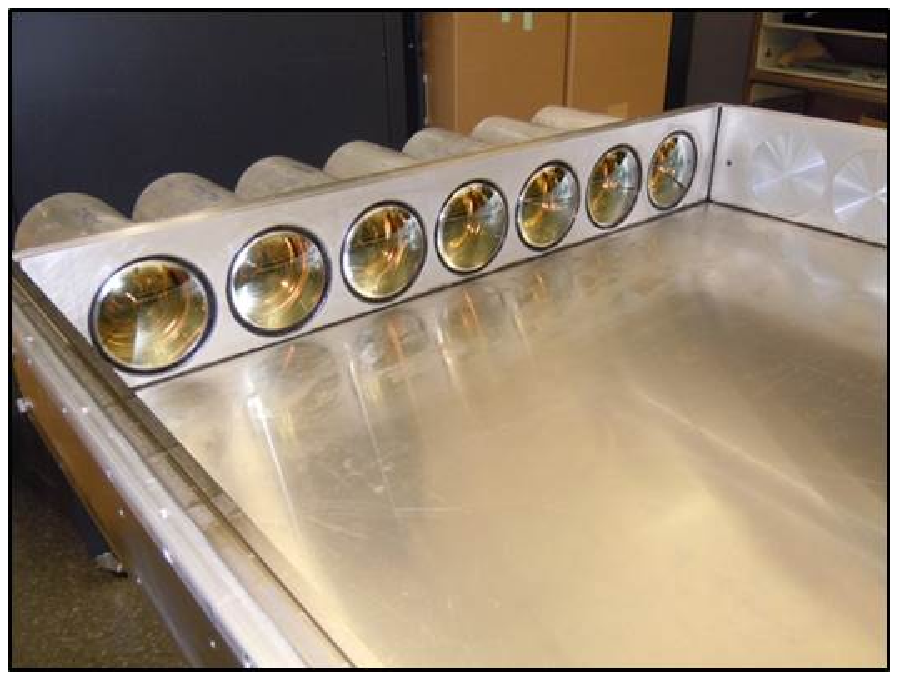} }
\subfigure[\label{aero-tray} Aerogel tray]
{\includegraphics[width=1.6in]{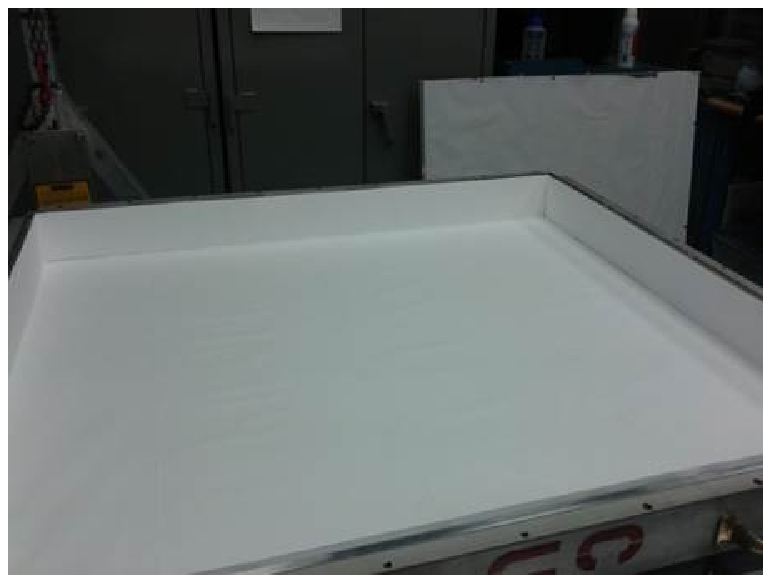}}
\caption{\label{fig-aero-tray-and-box} (Color online) Diffusion box (left, a) 
and aerogel tray (right, b) of the SHMS aerogel detector. Both are covered with
 a diffuse reflector material.}
\end{figure}

The SHMS aerogel detector is installed in the detector hut of the SHMS 
between the heavy gas \v{C}erenkov and the second hodoscope plane
(see Fig.~\ref{fig-aero-and-shms-hut}).  The detector consists of two main 
components: a tray which holds the aerogel material, and a light diffusion box 
with photomultiplier tubes (PMTs) for light readout, as illustrated in
Fig.~\ref{fig-aero-tray-and-box}. To cover the required kaon identification
momentum range of up to 7.2 GeV/$c$, four identical trays for aerogel of nominal
refractive indices of 1.030, 1.020, 1.015 and 1.011 were constructed.
The design allows for easy detector assembly and replacement of the aerogel
trays.

An active area of 60 cm (width) and 90 cm (height) fully satisfies
the needs of the approved experiments~\cite{E12-07-105, E12-09-011,
  E12-09-017, E12-06-104, E12-06-107} requiring kaon identification,
which will all use targets with lengths of about 10~cm. The momentum range
(around the chosen spectrometer central momentum) and angular phase space
accepted by the SHMS for an assumed 10 cm target length corresponds to en
assemble of particle rays in the SHMS detector hut, at the location of the
aerogel detector, of less than this active area. 
However, the detector was designed and built 
with a total area of 110 $\times$ 100 cm$^2$ to in principle also allow
coverage of the full momentum and angular range of the spectrometer for
a 40-cm long target.

Using 5-inch diameter PMTs mounted on the vertical sides of the diffusion box 
and up to 9 cm aerogel thickness in the trays, the total depth of the detector 
is 24.5 cm along the optical axis of the SHMS. Such a compact detector is well 
suited for the limited space available, up to 30 cm, in the SHMS detector stack.

Prior to construction, the performance requirements of the detectors were 
studied and optimized using a dedicated Monte Carlo code~\cite{doug98}. 
The baseline design for this optimization assumed a 
110$\times$100$\times$24.5 cm$^3$ detector box covered from the inside with a 
diffuse reflector of at least 96\% reflectivity, and with 5-inch diameter PMTs 
of 20\% quantum efficiency mounted on each long side of the diffusion box. 
This design concept follows earlier proven aerogel detector 
designs~\cite{Benot,alcorn04,Asaturyan}.  Since the photon detection
probability is directly proportional to the fraction of the inner
detector surface covered by the photo-cathode windows of the
PMTs~\cite{poelz86,carlson86}, the mechanically allowable maximum
number of seven PMTs on each side was assumed in these studies. 
The simulations showed that the number of photo-electrons $N_{pe}$ as
measured by all PMTs summed is uniform to within 10\% over the full active area
of the detector, with this two-sided readout.

The diffusion box is made of the aluminum alloy 6061-T6. The side panels
are constructed of $\sim$2.5 cm (1-inch) plates. The back cover is
$\sim$1.6 mm (1/16 inch) thick. The inner dimensions of the box are
$\sim 103\times 113\times 17.3~cm^3$ (40.5" $\times$ 44.5" $\times$6.82"). 
To optimize light collection the inner surface of the diffusion box is 
lined with either 3 mm (covering $\sim$60\% of the surface) or 1 mm (remaining
$\sim$40\% of the surface) thick GORE reflector material~\cite{Gore}. 
This material has a reflectivity of about 99\% over the entire spectrum.
Further discussion of this material and its effect on light
collection can be found in section~\ref{cosmic_tests}.

The light collection is handled by 5-inch diameter photomultiplier tubes.
The 5.56'' (14.1 cm) diameter cylindrical housings holding the
PMTs are mounted upon 14 waterjet cut circular openings on the left
and right (long) sides of the diffusion box, with minimum spacing of 14.92 cm 
(5.875'') between the centers. 
The PMTs are sealed into their housing using a light-tight synthetic rubber 
material (Momentive RTV103 Black Silicone Sealant) and the whole assembly is 
sealed light-tight. The mechanical design includes six openings on the top of 
the diffusion box, presently covered with blanks, that can be used to increase 
the signal output from the detector by about 30\%, if needed.

The magnetic shielding for the PMTs consists of 13.5 cm (5.316'') 
diameter $\mu$-metal cylinders, which were constructed to
end abreast with the PMT window.  The construction also features bucking
coils that can be installed on the PMTs, if excessive residual
magnetic fields appear to be present in the SHMS hut. 
The selection criteria for the PMTs and studies on efficiency of the 
magnetic shielding including the bucking coil studies, are presented in
section~\ref{magnetic-shield}.

The aerogel trays are of the same transverse size as the diffusion
box but 11.3 cm (4.45'') deep. The front cover of the trays is made of a
5~mm thick honeycomb panel with 
effective Aluminum thickness to $\sim$1.3 mm (0.050'').
The inner surface of the SP-30 and SP-20 aerogel trays is covered with 
0.45 $\mu$m thick Millipore paper Membrane GSWP-0010 (Millipore) of 
reflectivity of about 96\%~\cite{millipore}. 
Though Millipore is difficult to handle, its chemical inertness makes it 
superior to reflective paints. 
For the two lower refractive index trays (SP-15 and SP-11), in order to optimize
light collection, we used 1 mm thick Gore diffusive reflector material 
(DRP-1.0-12x30-PSA) with reflectivity of about 99\%.

For the \v{C}erenkov radiator high transparency aerogels were used. The higher 
two of the refractive indices (SP-30 and SP-20) were originally manufactured by
 Matsushita Electric Works, Ltd.  The lower two indices (SP-15 and SP-11) were 
manufactured by Japanese Fine Ceramics Center.  
These tiles have dimensions of approximately 11 cm by 11 cm by 1 cm.  
They feature a waterproof coating that make them
hydrophobic~\cite{adachi,Aschenauer}.  This removes the need for baking (which 
in fact would destroy the coating). Detailed studies of the aerogel 
characteristics are presented in section~\ref{aero_study}.

\begin{figure}[!htbp]
\centering
{
\includegraphics[width=1.6in]{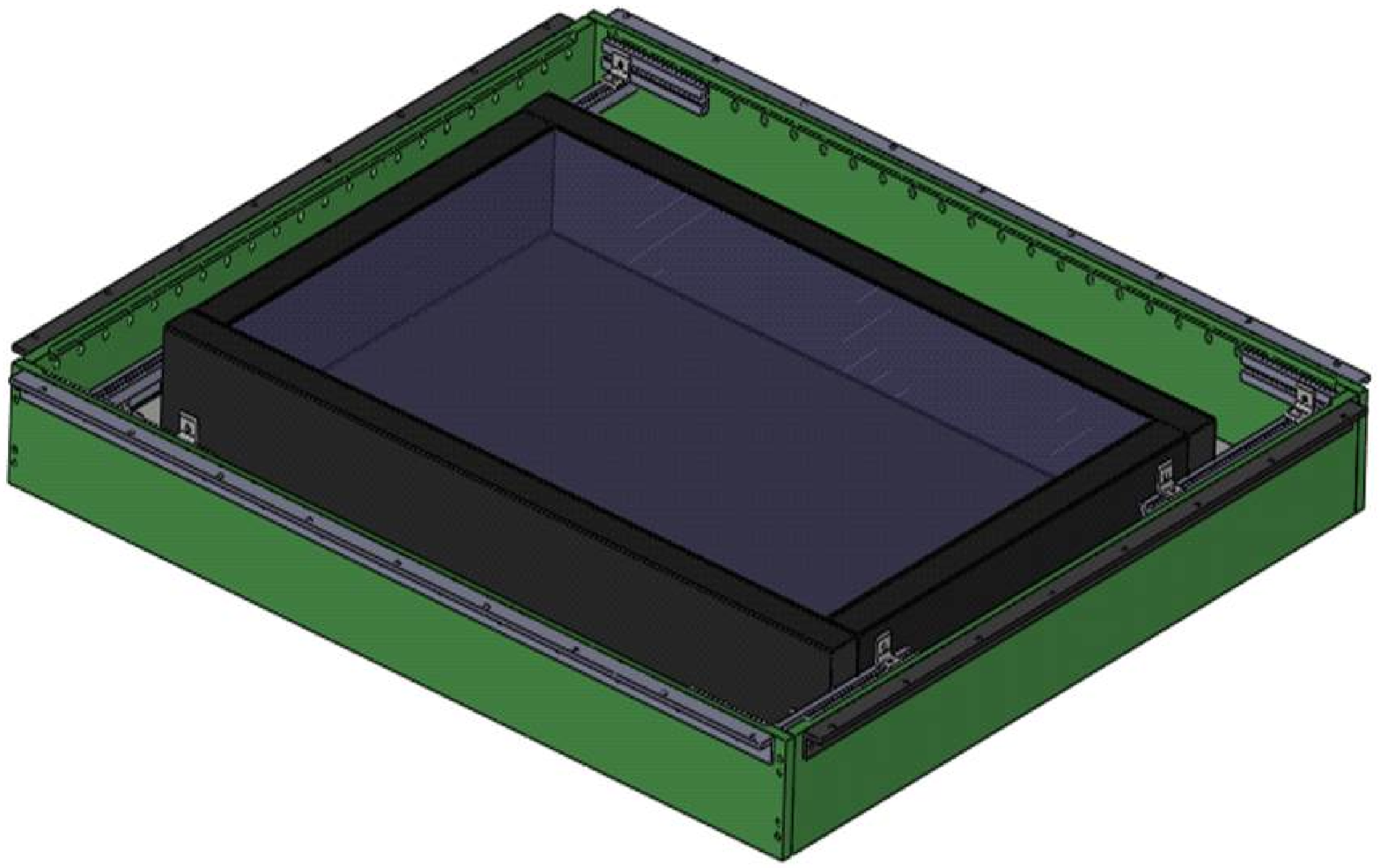} }
{\includegraphics[width=1.6in]{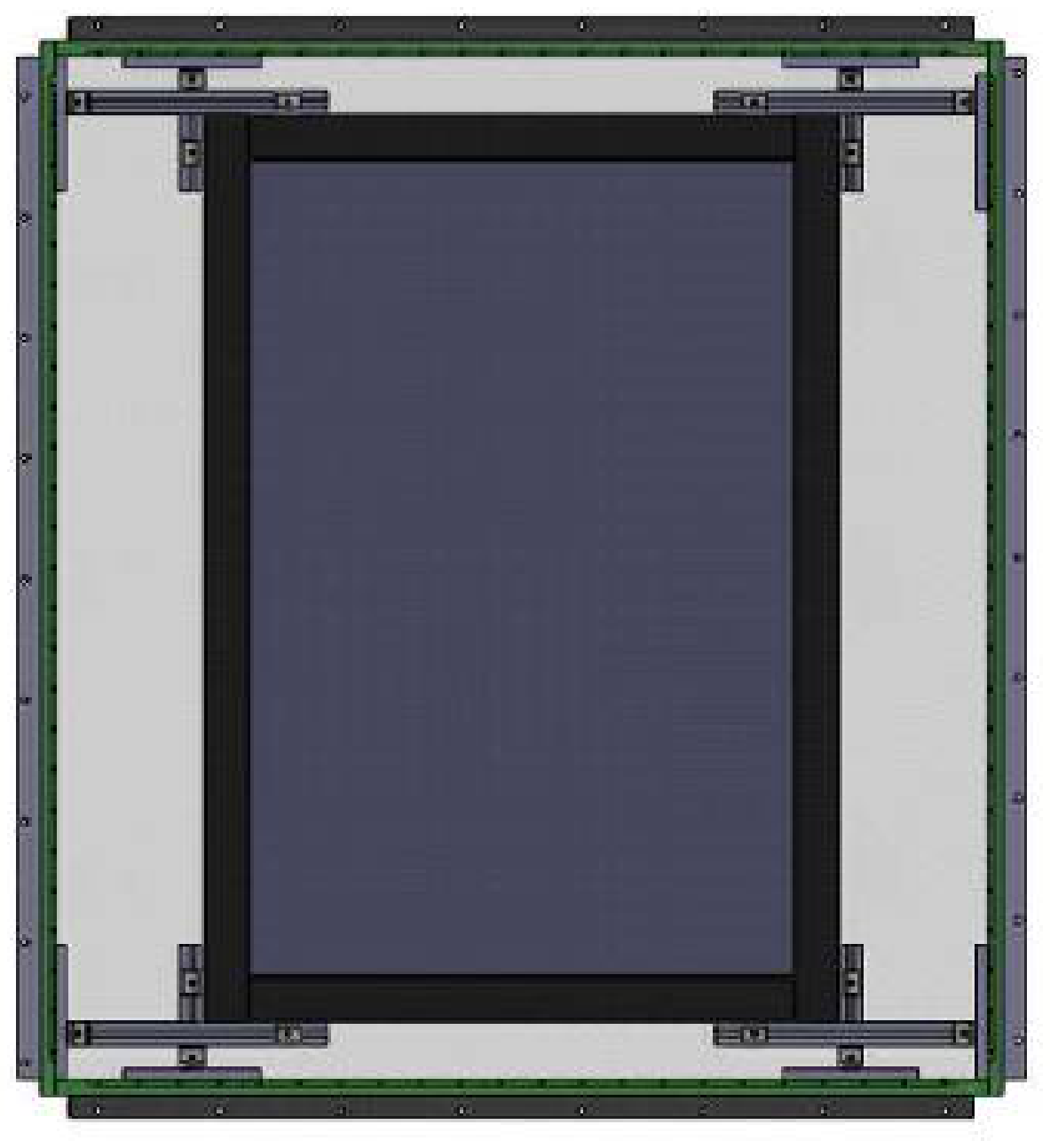}}
\caption{\label{fig-sp10-holder} (Color online) Schematic drawing of the SP-11 
aerogel tray with inner frame covering the active area of 90 cm x 60 cm. }
\end{figure}

The trays were filled with aerogel tiles layer by layer. In each layer
the tiles were layed down flat and arranged in a brick pattern to minimize
holes in the radiator. 
To fill gaps of less than the size of a full tile at the edges of the tray 
the aerogel material was cut using a diamond coated saw or razor depending on 
the refractive index of the material.
The aerogel radiator is on average $\sim$9 cm thick (8 layers).
The SP-30, SP-20 and SP-15 aerogel trays were filled over their entire 
110 cm x 100 cm area. The SP-11 aerogel tray radiator covers only the active
area of 90 cm x 60 cm required by the experiments. An inner frame has been 
designed to arrange the aerogel
tiles inside the active area of this tray (see Fig.~\ref{fig-sp10-holder}).
The sides of this inner frame are made of carbon fiber square tubes. This 
assembly allows future X-Y repositioning of the inner frame inside the tray. 

To protect the aerogel radiator from severe damage in case of accidental 
flipping over of a tray during installation, a net of thin stainless steel 
wires is installed in close proximity to the aerogel surface.
This is a technique previously tested in aerogel detectors at
JLab~\cite{Asaturyan}.  The wires form an interweaving grid by running
between stainless steel screws on the sides of the box. Small springs
attached to the ends of wires provide necessary tension.

An aerogel tray attaches to the diffusion box by means of bolting
through flanges surrounding both boxes. A round O-ring running in a
shallow groove along the diffusion box sides ensures a light tight connection.
The entire detector is designed so that it can be removed from the
sliding detector stand that positions the detector into the SHMS 
detector stack.

The SHMS aerogel detector was designed considering the optimization of the amount of material in the particles's path in the spectrometer. The amount of material across the sensitive area of the detector was minimized to reduce effects of multiple scattering and production of $\delta$-electrons, but with the optimal thickness of radiator required to achieve higher PID efficiency with the detector. Table~\ref{tab:radiationLength} summarizes the effective thickness of materials in the path of particles transversing the SHMS aerogel detector, with the respective radiation length of these materials.

\begin{center}
\begin{table}
\caption{\label{tab:radiationLength} List of materials in the path of particles traversing the SHMS aerogel detector, with their respective radiation length.}
{\centering  \begin{tabular}{|c|c|c|c|c|}
\hline
Component & Material & Thickness & Density & Radiation Length\\
          &          &  (cm) & $({\rm g}/{\rm cm}^3)$ & $({\rm g}/{\rm cm}^2)$\\
\hline
Window & Al & 0.13 & 2.7 & 24.01\\
Aerogel & SiO$_2$ & 9.0 & 0.2 & 44.054\\
Air gap & Air & 17.1 & 0.00121 & 36.66\\
Window & Al & 0.16 & 2.7 & 24.01\\
\hline
\end{tabular}\par}
\end{table}
\end{center}

\section{\label{aero_study} Aerogel characterization}

For the assembly of the trays with nominal refractive indices $n$=1.030 
and 1.020, aerogel tiles from an earlier experiment~\cite{Blast}
were used, where they had been exposed to a relatively harsh radiation 
environment. 
The trays with nominal refractive indices $n$=1.015 and 1.011 contain 
aerogel tiles acquired directly from the manufacturer albeit in
several batches. 
To check for possible degradation of the previously used tiles and 
dependence on the manufacturing process the properties and uniformity 
of the aerogel were carefully analyzed prior to the assembly of the 
detector. For each nominal refractive index a sample of 5 - 15\% of all the 
tiles was randomly selected and tested. In the following subsections, we 
report the studies of the most important properties of the aerogel for the 
construction of the detector: refractive index, tile dimensions, light yield,
light transmittance and absorption, and hydrophobicity.

\begin{figure}[!htbp]
\centering
\subfigure[\label{fig:ProtPic} Prototype picture] 
{\includegraphics[width=1.6in]{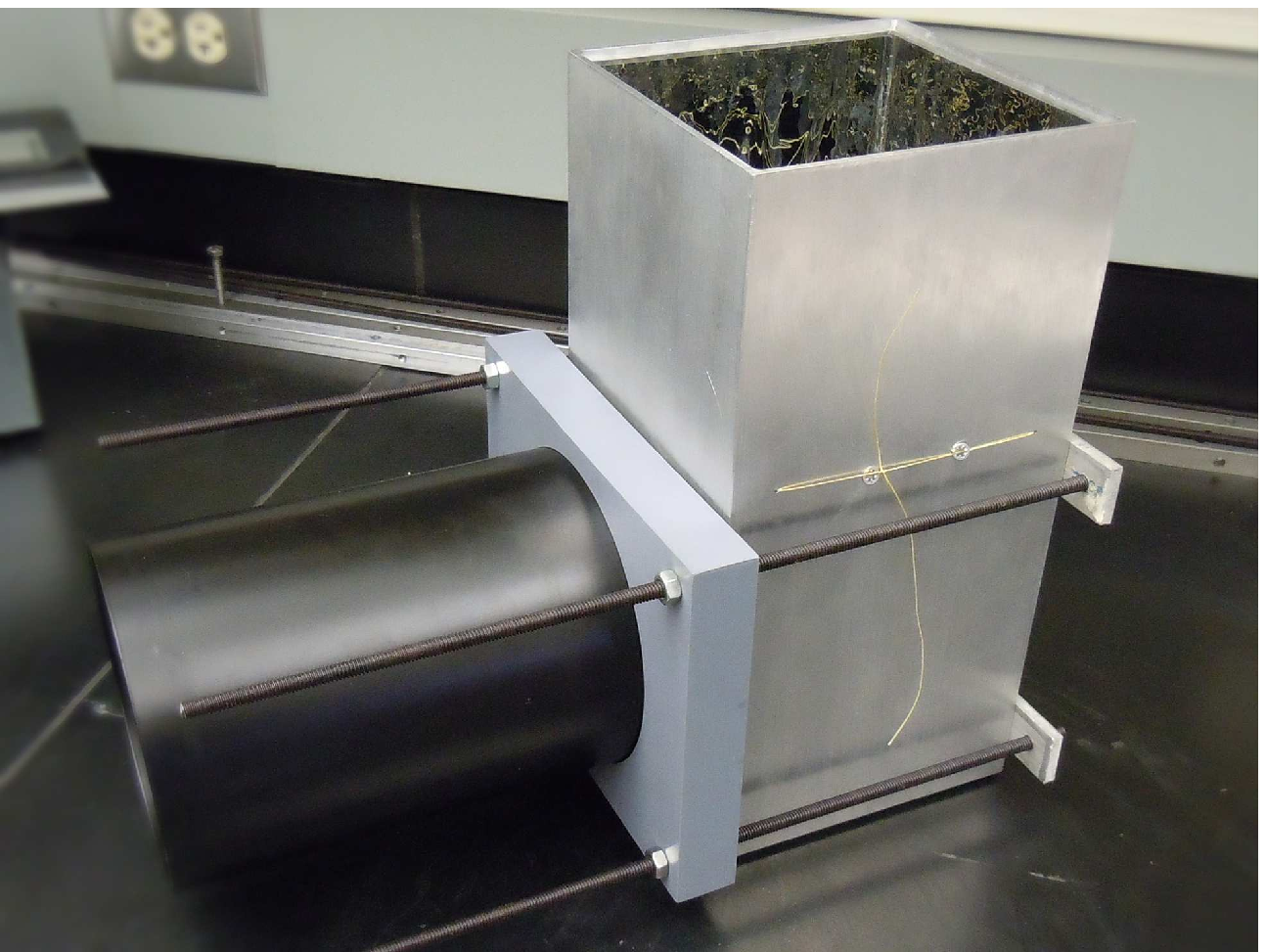} }
\hspace{0.1cm}
\subfigure[\label{fig:ProtDraw} Prototype drawing]
{\includegraphics[width=1.4in]{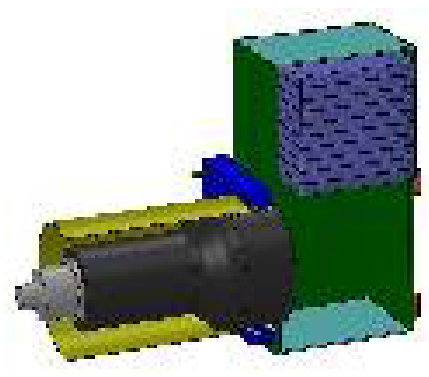} }
\caption{\label{fig:prototype} (Color online)  Prototype of the aerogel 
detector built for studies of the characteristics of the detector.
Left (a) the prototype with the top cover of the aerogel box removed.
Right (b) a schematic drawing of the prototype showing the position of the PMT
(in gray) and aerogel tiles (blue) stacked inside the prototype.}
\end{figure}

Several features of the detector like the selection of reflector material 
or thickness of the aerogel radiator also required careful studies before 
the final assembly. A prototype counter was constructed to study and 
optimize these features of the actual detector.
The prototype was a combination of a small aerogel tray (which can be filled 
with up to 14 aerogel tiles) and a diffusion box viewed by a single PMT.
The internal surface of the prototype was covered with Millipore
material or aluminized Mylar as reflective material.
The prototype is shown in Fig.~\ref{fig:prototype}. 

This single-PMT prototype allowed studies of, {\sl e.g.}, different reflective 
materials covering the walls of the detector, comparative light yield of the 
different aerogel refractive indices, and optimization of performance with PMTs.
After the conclusion of the tests, the construction of the entire array of the 
actual detector proceeded without significant changes from the prototype.

\subsection{\label{subsec:aero_refracIndex} Refractive index}
The velocity threshold for production of \v{C}erenkov radiation inside a 
material is given by $v_t=c/n$, where $c$ is the speed of light in vacuum and 
$n$ is the refractive index of the material. 
The momentum threshold of a particle associated with $v_t$ is given by 
$p_t = m/\sqrt{(n^2-1)}$, where {\it m} is the mass of the particle.

\begin{figure}
\begin{center}
\epsfxsize=3.40in
\epsfysize=3.00in
\epsffile{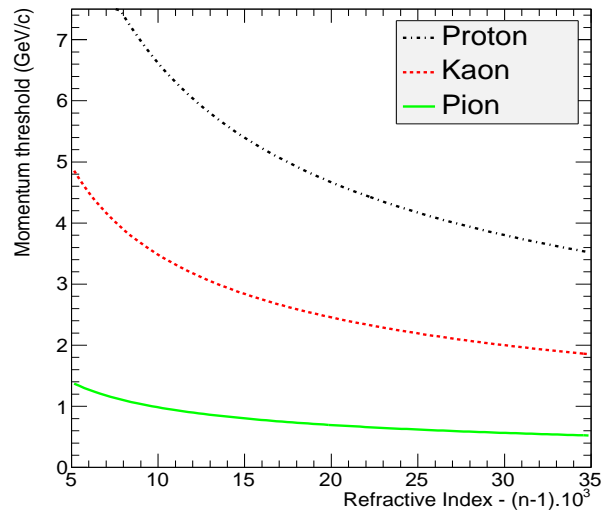}
\caption{\label{fig:aero_momentumThreshold} (Color online) Momentum threshold 
for proton, pion and kaon above which \v{C}erenkov light is radiated in 
aerogel, according to the aerogel refractive index. On this horizontal scale, 
SP-11 (SP-15, SP-20, SP-30) corresponds to 11 (15, 20, 30, respectively). }
\end{center}
\end{figure}

In the SHMS Aerogel \v{C}erenkov Detector, the refractive index of each tray 
of aerogel defines the momentum range that the detector will allow for 
distinguishing between kaons and protons in the PID system of the SHMS. 
Fig.~\ref{fig:aero_momentumThreshold} shows the dependence of the momentum 
threshold, $p_t$, for radiation of \v{C}erenkov light for pion, kaon and 
proton as a function of the refractive index of the aerogel.

\begin{figure}[H]
\begin{center}
\epsfxsize=2.27in
\epsfysize=2.00in
\epsffile{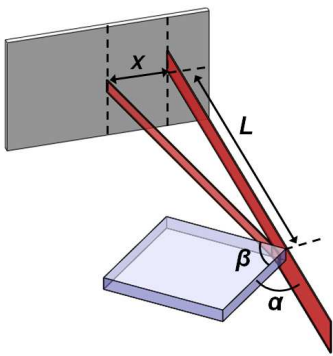}
\caption{\label{fig:aero_RefracIndex_scheme} (Color online) Schematic of the 
setup used to measure the refractive index of aerogel tiles. A beam of light 
is incident over one edge of the tile. Refraction of the outgoing light is 
used to calculate its refractive index based on Snell's Law.}
\end{center}
\end{figure}

Measurements of the refractive index of tiles were performed with a technique
based on Snell's Law. A similar method was used as in Ref.~\cite{buzykaev},
it consists of measuring the refraction of a beam of light (red, 
wavelength of {\it 670~nm}) incident on one corner of an aerogel tile.
As illustrated in Fig.~\ref{fig:aero_RefracIndex_scheme}, the beam of light is 
incident on the tile at an angle $\alpha$ with respect to the normal to the 
surface of its corner. 
The section of the beam of light that goes through the analyzed tile is 
refracted and hits a screen placed at a distance {\it L} from the tile. 
Refraction bends the incident beam of light in the aerogel material, so that it
 is separated from the non-refracted part of the beam that does not go through 
the aerogel (direct beam) by a distance {\it x} on the screen. 
From these variables and the angle $\beta$ between two sides of the aerogel 
tile (90~degrees in our case), one can determine the refractive 
index {\it n} of the tile using,
\begin{equation}
\frac{n}{n_{air}} = \sqrt{\frac{sin^2(\alpha)+sin^2(\gamma)}{sin^2(\beta)}+
2\frac{sin(\alpha)\sin(\gamma)}{tan(\beta).sin(\beta)}} ,
\label{eq:RefractiveIndexMethod}
\end{equation} 
where
\begin{equation*}
\gamma = tan^{-1}\left(\frac{x}{L}\right) - \alpha + \beta ,
\label{eq:refrIndexMethod_gamma}
\end{equation*}
and $n_{air}$ is the refractive index of the air surrounding the aerogel tile. 
Based on the temperature, atmospheric pressure and relative humidity of our 
laboratory, we considered $(n_{air}-1)\times10^5$~=~26.5~$\pm$~0.5, 
according to \cite{NIST_air_refIndex}.

\begin{figure}
\begin{center}
\epsfxsize=3.40in
\epsfysize=3.00in
\epsffile{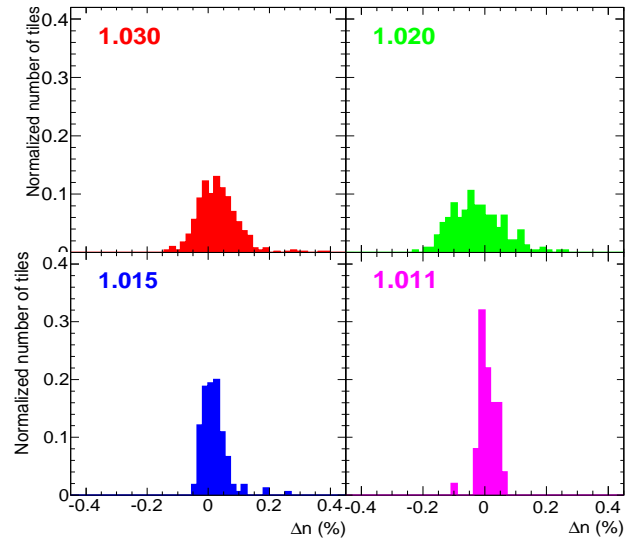}
\caption{\label{fig:refrIndex} (Color online) The deviation of the refractive 
index of the individual tiles to the nominal refractive index of each tray. 
The distributions were measured for a sample of 5-10\% of the tiles used for 
the assembly of the four trays of the detector.}
\end{center}
\end{figure}

Figure~\ref{fig:refrIndex} shows the statistical 
distribution of the refractive index measured for a randomly selected sample of
 the aerogel tiles used in the assembly of the trays. 
The dominant source of systematic uncertainties in this setup is the 
measurement of the angles $\alpha$ and $\beta$. 
The angles can have an offset of up to $0.8$~degrees resulting in a 
systematic error of $\pm~0.72\times10^{-3}$ on the refractive index.

The SP-30 tiles of nominal index $n$=1.030 had a measured mean value of 
$n$=1.0303 and a standard deviation of 0.0007. 
For the SP-20 tiles of nominal index $n$=1.020, the measured mean value was 
$n$=1.0198 with a standard deviation of 0.0009. 
The SP-15 tiles had a mean value of $n$=1.0152 and a standard deviation of
0.0004.
Finally, for the SP-11 tiles, the mean value was $n$=1.0111 with a standard 
deviation of 0.0003.
The uncertainty of the refractive index of air is a decade less, thus the 
systematic errors are dominated by the statistical variation of the refractive 
index from tile to tile. The refractive indices and uncertainties are 
summarized in Table~\ref{tab:aero-refractiveIndex}.

We do note that the manufacturer of the aerogel reported the refractive index 
of the tiles used in the tray SP-15 and SP-11 as $n$=1.01495~$\pm$~0.00020 and 
$n$=1.01066~$\pm$~0.00010, respectively. The ratio of these numbers
and the mean values we measured is about 1.0003, similar to $n_{air}$ and, for 
this reason, we believe the manufacturer may have calculated their values of 
refractive index with respect to air, whereas we calculated the refractive index
of aerogel relative to vacuum.

\begin{table}
\begin{center}
\caption{Aerogel refractive index measured for a sample of the tiles used on 
the construction of each tray. The refractive index of aerogel was calculated 
using equation~\ref{eq:RefractiveIndexMethod} assuming $n_{air}$=~1.000265.}
\label{tab:aero-refractiveIndex} \smallskip
\small
\begin{tabular}{c c}
\hline
Tray & Refractive index \\
\hline
1.030 & 1.0303 $\pm$ 0.0007\\
1.020 & 1.0198 $\pm$ 0.0009\\
1.015 & 1.0152 $\pm$ 0.0004\\
1.011 & 1.0111 $\pm$ 0.0003\\
\hline
\hline\end{tabular}
\end{center}
\end{table}

\subsection{\label{subsec:aero_dimensions} Dimensions of the aerogel tiles}

The thickness and the width of a sample of aerogel tiles were measured.  
For the measurement of the aerogel tile thickness each aerogel tile was 
placed in between two aluminum plates. 
Using a caliper, we measured the thickness of each tile from the distance 
between these two aluminum plates. Similar, a caliper was used to measure 
the lateral width of each tile of the selected sample of aerogel. The distance 
of two opposite sides of the tiles was measured in several points along the 
dimension of each aerogel tile and an average over these measurements was 
considered as the tile width.
Table~\ref{tab:aero-dim} summarizes the dimensions of the aerogel tiles used in
 the assembly of the SHMS Aerogel \v{C}erenkov Detector.

\begin{table}[H]
\begin{center}
\caption{Dimensions of (a sample of) aerogel tiles used on the construction of 
the detector's trays.}
\label{tab:aero-dim} \smallskip
\small
\begin{tabular}{c c c}
\hline
Tray & Width (mm) & Thickness (mm) \\
\hline
1.030 & 113.10~$\pm$~0.40 & 11.583~$\pm$~0.067 \\
1.020 & 110.82~$\pm$~0.59 & 11.42~$\pm$~0.33 \\
1.015 & 111.83~$\pm$~0.22 & 11.10~$\pm$~0.15\\
1.011 & 112.28~$\pm$~0.35 & 10.93~$\pm$~0.10\\
\hline
\hline\end{tabular}
\end{center}
\end{table}

\subsection{\label{subsec:aero_LightYield}Light yield - \v{C}erenkov radiation}

The basic physical principle behind the particle identification with the SHMS 
Aerogel \v{C}erenkov detector is the emission of \v{C}erenkov radiation in the 
aerogel when the speed of a particle going through the detector is larger than 
the phase velocity of the electromagnetic fields inside this material, 
{\it i.e.} $v>c/n$. 
Combined with knowledge on the momentum of the detected particles as provided 
by the magneto-optics analysis of the events in the spectrometer,
the production of \v{C}erenkov radiation is then used as a mass analyzer.

Based on classical electrodynamics and the quantization of light, one
can estimate the energy ($E$) spectrum and the number of photons 
({\it N}) produced per unit track length ({\it x}) of a particle with charge 
$ze$ and per unit energy interval of the photons according to 
Ref.~\cite{Jackson} as 

\begin{equation}
\frac{d^2N}{dE{}dx} = 
\frac{\alpha{}z^2}{\hbar{}c}\left(1-\frac{1}{{\beta^2}n^2(E)}\right) ,
\label{eq:Ncher_model1}
\end{equation}
or, equivalently,

\begin{equation}
\frac{d^2N}{dx{}d\lambda} = 
\frac{2\pi \alpha{}z^2}{\lambda^2}\left(1-\frac{1}{{\beta^2}n^2(\lambda)}\right) ,
\label{eq:Ncher_model2}
\end{equation}
where $\alpha$ is the fine structure constant, $\hbar$ is the Planck's 
constant, {\it c} is the speed of light, and $\beta$ is the fractional 
velocity of the incoming particle ($\beta = v/c$). 
{\it n} is the refractive index of aerogel and a function of photon energy
(or wavelength $\lambda$).

\begin{figure}[H]
\begin{center}
\epsfxsize=3.20in
\epsfysize=3.20in
\epsffile{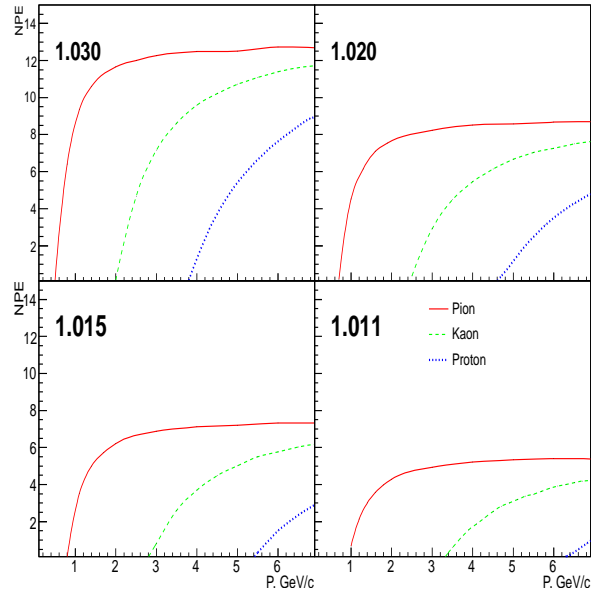}
\caption{\label{fig-shms-npe} (Color online) Estimated mean number of 
photoelectrons (NPE) to be measured by the ``baseline design'' SHMS Aerogel
\v{C}erenkov Detector for the trays with different refractive index. 
This calculation was done  following the model proposed by the authors of 
Ref.~\cite{doug98}, for the simulation of an aerogel \v{C}erenkov similar to 
our detector. }
\end{center}
\end{figure}

Figure~\ref{fig-shms-npe} shows an estimate of the total light yield 
expected for the different nominal refractive indices of aerogel, for a 
thickness of 10 cm.
This estimate is based on the method discussed in Ref.~\cite{doug98}, where 
effects such as light absorption in the components of the detector and
quantum efficiency of the PMTs were taken into account. Note that this estimate
is done for the baseline design aerogel \v{C}erenkov detector as described in
section~\ref{design}.

The calculated number of photo-electrons in saturation (for $\beta=v/c\sim$1 
particles), is $\sim$12 for aerogel with nominal refractive index $n$=1.030, 
and $\sim$4.6 for $n$=1.011.  
Due to their relatively lower velocity ($\beta<$1) kaons and protons in the 
same momentum range are far from saturation.
Figure~\ref{fig-shms-npe} shows the calculated numbers of photo-electrons for 
kaons and protons over the entire momentum range required by the experiments 
in section~\ref{intro}.

\begin{figure}[H]
\centering
{\includegraphics[width=\columnwidth,height=2.5in]{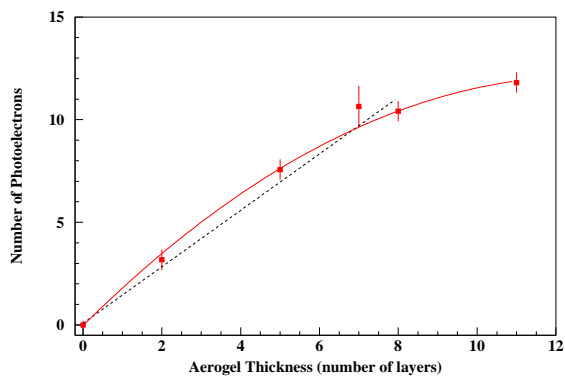}}
\vspace{-0.5in}
\caption{\label{fig:aero-quality} (Color online) The number of photoelectrons 
versus the aerogel radiator thickness (varying the number of layers) generated
by cosmic muons in the single counter prototype. 
The dotted line represents a linear fit to the experimental data up to an 
aerogel thickness of 8 layers ($\sim$9 cm). 
The solid curve shows a second-order polynomial fit to all data set up to a 
thickness of 11 layers ($\sim$12 cm). Aerogel tiles with refractive index 
$n$=1.030 were used.}
\end{figure}

To check the relative optical quality of the aerogel material we have measured 
the light yield generated by cosmic muons using the single counter (prototype) 
described above. Fig.~\ref{fig:aero-quality} shows the number of photoelectrons
as a function of aerogel thickness (or number of layers). 
We measure a nearly linear dependence of the number of photoelectrons on the 
aerogel thickness up to $\sim$9~cm (8 layers), as illustrated by the dotted 
line in Fig.~\ref{fig:aero-quality}, although some saturation may set in around
 6 layers or so. 
For aerogel thicknesses of more than 11~cm  (10 layers) the light yield clearly
 saturates as shown by the second-order polynomial fit (solid curve) - some of 
the produced \v{C}erenkov light gets reabsorbed.

To check for possible degradation of previously used aerogel material, we 
compared the results of measurements using 10 tiles ($\sim$ 11 cm) of new and 
previously used $n$=1.030 aerogel. Both aerogel material having seen the 
radiation dose of the previous experiments and new aerogel material are 
indistinguishable.
These tests suggest, within the uncertainty of the measurements, that the 
previously used aerogel tiles have a light yield similar to the new tiles and 
do not show significant aging or radiation degradation. 
Our result is consistent with the investigations of radiation damage of  
aerogel material in Ref.~\cite{Bellunato,Sumiyoshi}. 
There, aerogel tiles were exposed to very intense $\gamma$-radiation from 
a $^{60}$Co source, and to proton and neutron high intensity beams. 
Transmittance, clarity factor and refractive index of the aerogel tiles were 
measured before and after irradiation and no detectable degradation of the 
optical parameters was observed up to doses of $\sim$10
Mrad. Ref.~\cite{Sumiyoshi} also showed that cracks in the aerogel tiles 
do not make a noticeable difference in the light yield.

The same selection criteria on optical properties and geometrical conditions 
(no chipped edges) were thus used for all tiles in the assembly of the SHMS 
Aerogel \v{C}erenkov Detector.

\subsection{\label{subsec:aero_optics} Aerogel Light transmittance}

To quantify the transmittance of light through aerogel, a 
Perkin/Elmer LAMBDA 750 UV/Vis/NIR Spectrophotometer was used to measure the 
light transmittance for a randomly selected sample of aerogel tiles for each 
nominal refractive index.
In this spectrometer, a beam of light with tunable wavelength, $\lambda$, is 
split into two beams. The first beam goes directly to the light sensor for the 
measurement of the reference light intensity, while the second beam goes through one 
aerogel tile. The thickness of aerogel that the second beam goes through is 
shown in Table~\ref{tab:aero-dim}.

The light transmittance of an aerogel tile is measured from the intensity of 
light that does not scatter or that is not absorbed by the aerogel, measured 
relative to the intensity of the reference beam.

\begin{figure}[H]
\begin{center}
\epsfxsize=3.40in
\epsfysize=3.00in
\epsffile{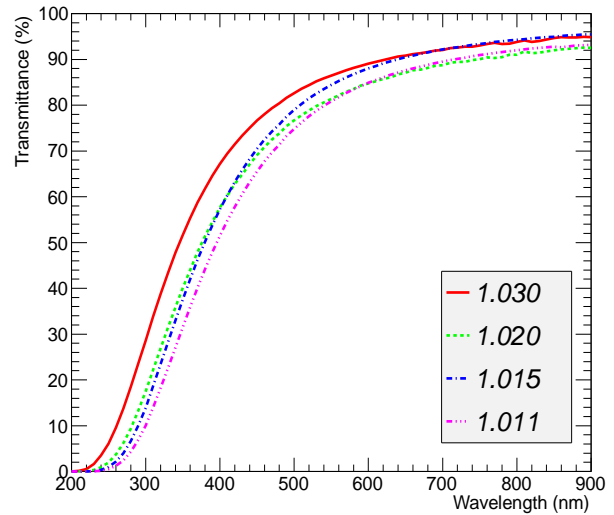}
\caption{\label{fig:aero_transmittance} (Color online) Mean value of the light 
transmittance of aerogel measured over a sample of $\sim10\%$ of the tiles used 
on the construction of the detector. The mean thickness of the tiles for each 
refractive index is shown in Table~\ref{tab:aero-dim}.}
\end{center}
\end{figure}

Fig.~\ref{fig:aero_transmittance} shows the mean values of the aerogel's 
transmittance for wavelengths ranging from 200~nm to 900~nm as analyzed by the 
spectrometer.
SP-30 tiles have a higher transmittance than SP-20, SP-15 and SP-11 tiles for 
wavelengths less than 600 nm including the region of interest for the SHMS 
Aerogel \v{C}erenkov Detector.
In this region the transmittance of SP-20 and SP-15 is similar and SP-11 has 
the lowest transmittance. For wavelengths greater than 600 nm the 
transmittance of the SP-15 tiles becomes similar to that of SP-30 and even 
exceeds it for wavelengths of greater than 700 nm.

The transmittances for different tiles of the same refractive index are tightly
clustered. We found that the statistical fluctuation of light transmittance 
for the different tiles with the same refractive index is less than 4\%. 
The systematic uncertainty of the measurements is $\pm$0.1\%.

\subsection{\label{subsec:aero_absorption} Aerogel light absorption}

Aerogel \v{C}erenkov detectors can exhibit light loss due to scattering and
absorption in the aerogel material, as determined by the scattering and
absorption legths.

\begin{figure}[H]
\centering
\subfigure[\label{fig:intSphere_trans} Transmittance mode.] 
{\includegraphics[width=1.6in]{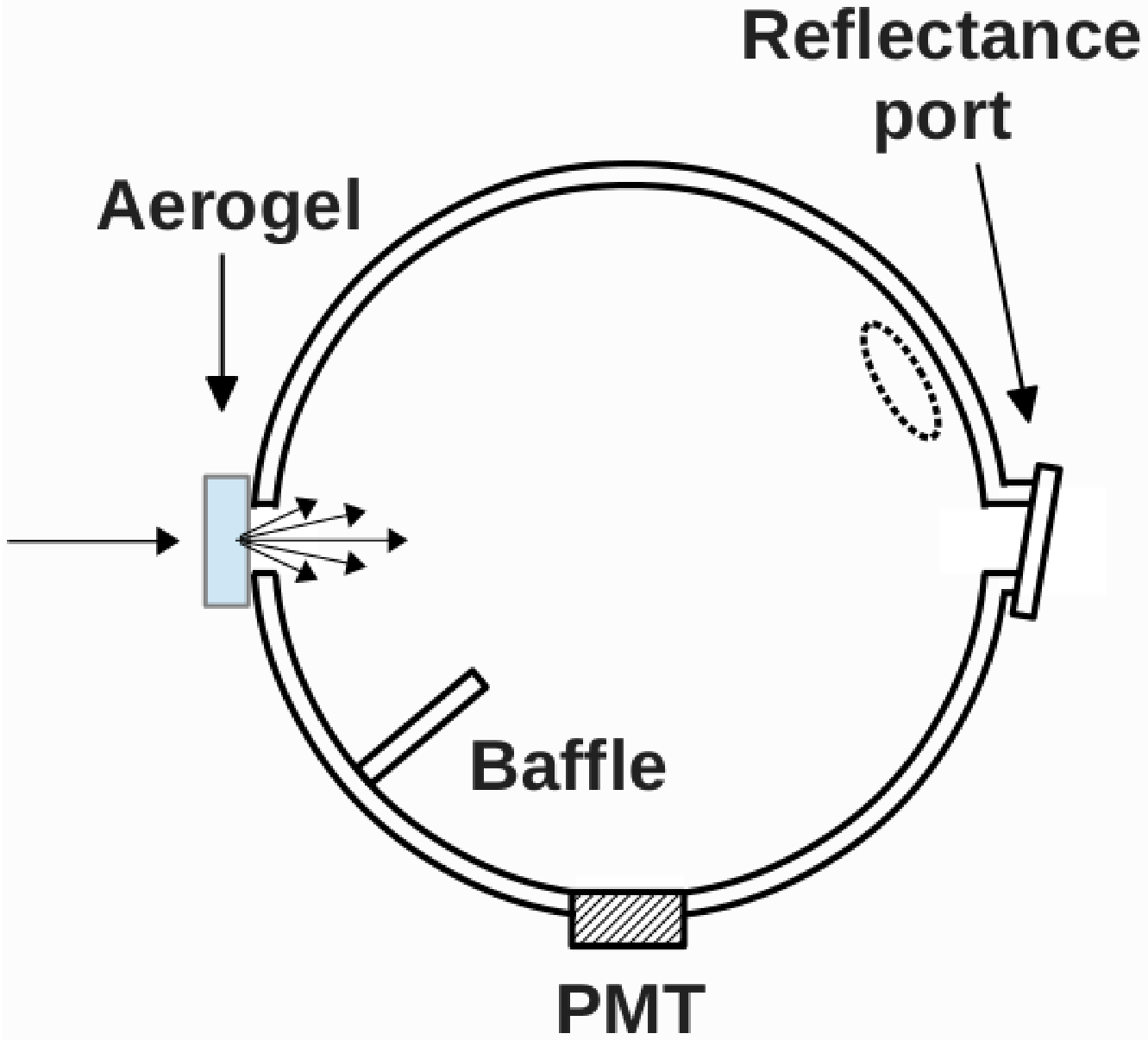} }
\hspace{0.1cm}
\subfigure[\label{fig:intSphere_scatt} Backscattering mode with forward light trap]
{\includegraphics[width=1.4in]{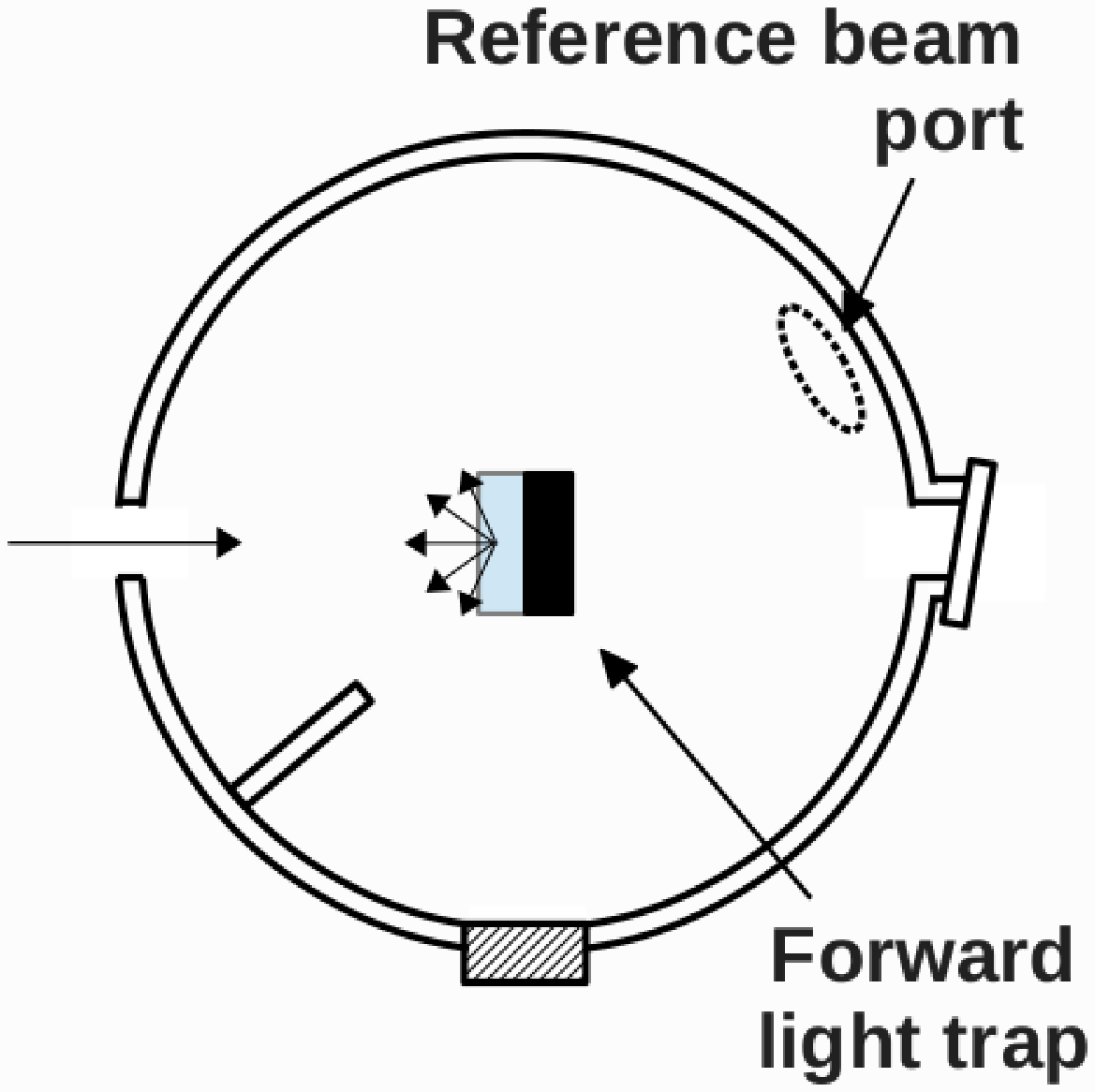} }
\caption{\label{fig:intSphere} (Color online)  The integrating sphere equipped for transmittance and backscattering measurements.}
\end{figure}

\begin{figure}[H]
\begin{center}
\epsfxsize=3.40in
\epsfysize=3.00in
\epsffile{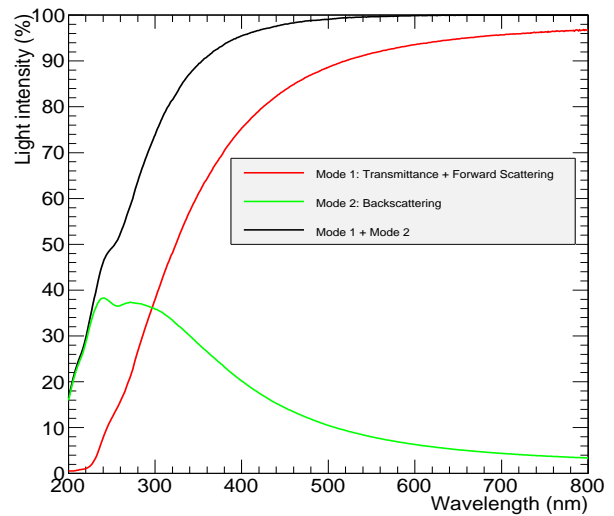}
\caption{\label{fig:aero_sp30_absorption} (Color online) Typical measured transmittance and backscattering for a range of wavelengths ranging from 200 to 800 nm in steps of 1 nm.}
\end{center}
\end{figure}

We measured the light absorption in the aerogel tiles using a Perkin-Elmer Lambda~950 spectrophotometer with a 150 mm PbS integrating sphere. The latter is covered with Spectralon, a material with high reflectivity. The integrating sphere can be operated in transmittance or backscattering mode as illustrated in Fig.~\ref{fig:intSphere}. The transmittance mode measures the light intensity that is either transmitted directly through the aerogel or scattered in the forward direction. The backscattering mode takes advantage of a forward light trap, which effectively allows one to measure only backscattered light from the aerogel. To determine the light absorption in aerogel, measurements were taken with the transmittance mode, with the sample outside the integrating sphere, and with the backscattering mode, where the sample was installed on a center mount with forward light trap inside the sphere. To take into account possible backgrounds we also took data without the aerogel installed in backscattering mode. Fig.~\ref{fig:aero_sp30_absorption} shows the measured light intensity of aerogel of refractive index $n$=1.030 in transmittance and backscattering modes. 

The sum of the light intensity - $I(\lambda)$ - measured in transmittance and backscattering modes are related to the absorption length of the aerogel through the Beer-Lambert law,

\begin{equation}
I(\lambda)=e^{-t/\Lambda_{abs}(\lambda))},
\label{eq:aero-trans-absorp-scatter}
\end{equation}

\noindent where $t$ is the thickness of the analyzed sample.

\begin{figure}[H]
\begin{center}
\epsfxsize=3.40in
\epsfysize=3.00in
\epsffile{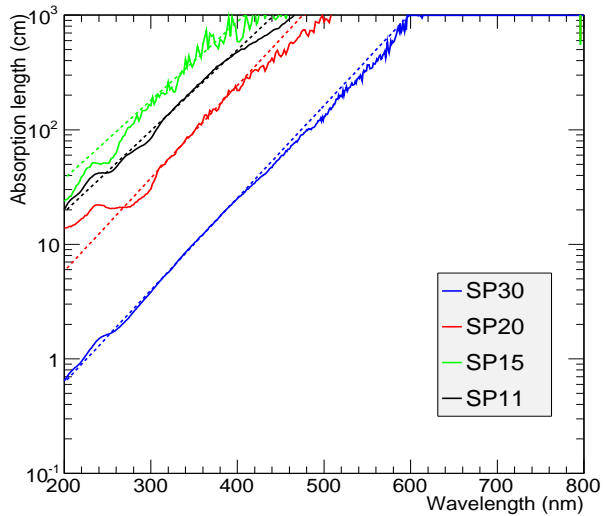}
\caption{\label{fig:aero_spall_absCoeff} (Color online) The absorption length determined from the forward and backward scattering measurements, e.g., those presented in Fig.~\ref{fig:aero_sp30_absorption} for SP-30 aerogel, using Eq.~\ref{eq:aero-trans-absorp-scatter}.}
\end{center}
\end{figure}

Fig.~\ref{fig:aero_spall_absCoeff} shows the absorption length for all four refractive aerogel indices calculated from Eq.~\ref{eq:aero-trans-absorp-scatter} and subtracting the background from the empty light trap measurement. The absorption length increases exponentially for wave lengths ranging from 200 to 400-600 nm depending on the refractive index. More light is absorbed for the highest refractive index aerogel. For example, at 400 nm the absorption length for SP-30 aerogel is 20 cm. This value is consistent in magnitude with that reported in Refs.~\cite{Aschenauer,buzykaev}. At about 450 nm, where our PMTs are most sensitive, the absorption length of SP-30 aerogel is 90 cm. The absorption lengths of SP-20, SP-15, and SP-11 aerogels in the range 400-450 nm are larger, on the order of a few hundred cm. The measured absorption lengths should also be compared to the total thickness of the aerogel radiator used in the detector, which is 10 cm. This suggests that the effect of absorption in the wavelength range of PMT sensitivity (380-450~nm) is negligible for all refractive aerogel indices, and even for the worst case, the SP-30 aerogel, the major fraction of the radiated light not be absorved in the aerogel. We note that in the region above 400-600 nm depending on the refractive index we only report the low limit for the light absorption length.

Earlier studies show that for the variety of aerogels the absorption length is nearly constant between 320 and 900 nm, and drops down logarithmically in the range 320-200 nm \cite{Aschenauer,buzykaev,Danilyuk,Barnyakov}. Typical values of the absorption length for aerogels with refractive indexes $n$=1.010 - 1.060 at $\lambda\geq$400 nm are reported as $>$40-60 cm, consistent with the values we measured.

\subsection{\label{subsec:aero_hydrophobicity} Aerogel hydrophobicity}

The density of the aerogel (directly related to its refractive index) can be 
significantly changed if the aerogel absorbs the humidity from the 
surrounding environment.
To eliminate this effect, some aerogel manufacturers use a hydrophobic coating 
on the tiles. This makes it possible to use the aerogel material in detectors 
for a long period of time without requiring special flushing with dry gas or 
any periodic maintenance.

The aerogel tiles used for the construction of the SHMS Aerogel \v{C}erenkov 
detector feature such a hydrophilic coating. 
To check the hydrophobicity of the aerogel, tests were carried out under 
severe humidity conditions. First, we kept some samples of the tiles for 24 
hours at the relative humidity 84$\pm$2\%, and later at 91$\pm$2\%. 
No significant change of the aerogel optical properties, e.g., the refractive 
index, were observed: the average change in refractive index due to humidity 
was measured to be less than +0.00010.

\begin{figure}
\centering
{\includegraphics[width=3.0in]{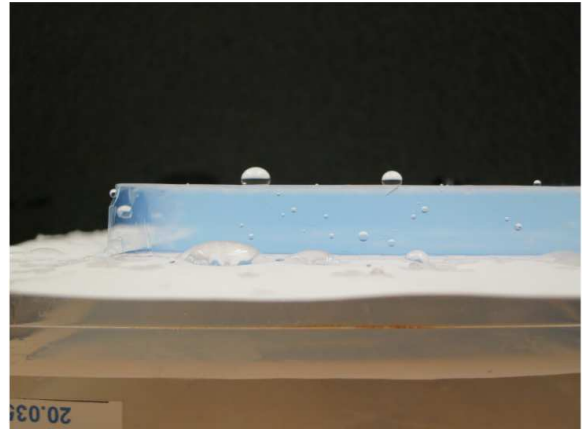}}
\caption{\label{fig:aero-hydrophob} (Color online) Drops of water on the 
hydrophobic surface of an aerogel tile. } 
\end{figure}

Finally, drops of water were added directly on the aerogel surface for a 
``wettability test'' (as shown in Fig.~\ref{fig:aero-hydrophob}), in which the 
contact angle of the water droplets to the aerogel surface is measured.
For a hydrophilic surface the contact angle is expected to be less than 90 
degrees, while hydrophobic surfaces have contact angles of more than 90
degrees. Our tests confirm that the aerogel tiles used in the SHMS Aerogel 
\v{C}erenkov Detector are hydrophobic.

\section{PHOTOMULTIPLIER TUBES STUDIES}
\label{pmt_study}

The main goal of these studies was to select the best PMTs for the SHMS aerogel
 detector from a pool of $\sim$70 PMTs model XP4500B previously used in the BLAST 
experiment~\cite{Blast} and from $\sim$25 PMTs model XP4572B previously used in the G0 
experiment~\cite{G0}.  
All of these are Photonis 10-stage PMTs, with semi-transparent bi-alkali 
photocathode with peak quantum efficiency $\sim20\%$ at wavelength
$\lambda\sim$350-450 nm, and gain of $\sim2\times 10^7$ at an operating high 
voltage 1800-2100 V.  
Their internal structure is similar, but the XP4572 photocathode is flat,
while the XP4500 photocatode is semi-spherical.

The main requirements for PMT selection were high quantum efficiency, 
low noise, high gain at relatively low high voltage (HV), and good single 
photoelectron resolution. 
For all PMTs the gain and its dependence on high voltage, and relative 
quantum efficiency were measured. For randomly selected PMTs we also 
studied photocathode uniformity, and effect of external magnetic 
fields and shielding on the PMT performance.
The final selection of PMTs for the detector consists of 14 PMTs of type XP4572
 based on the results of the studies described in the following subsections.


\subsection{Measurements of PMT gain and relative quantum efficiency}
\label{pmt_gain_qe}

\begin{figure}
\begin{center}
\epsfxsize=3.50in
\epsfysize=2.50in
\epsffile{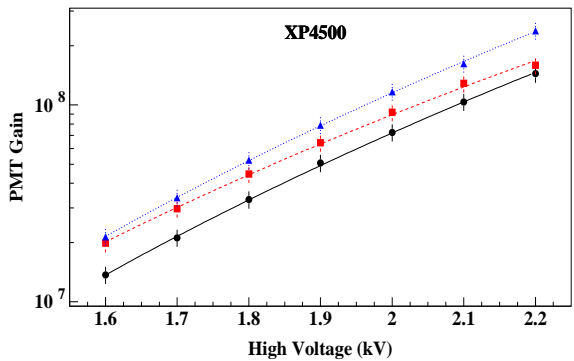}
\vspace{-0.5in}
\caption{\label{xp4500-gain} (Color online) Gain versus high voltage for three
  Photonis XP4500 PMTs. The data are fitted with a function $G \approx
  \alpha \times x^{\beta\times n}$, where $n$=10 is the number of dynodes,
  and $\alpha$ and $\beta$ are free parameters. 
At an operating high voltage of 1.7-1.8 kV the gain of the 
XP4500 PMT is on average $\sim3\times10^7$. }
\end{center}
\end{figure}

At any operating HV, gain measurements typically begin with localizing the 
single-electron peak (SEP).  The standard method to measure the SEP is to shine
a  light source on the photocathode and analyzing the resulting spectrum. 
Here, we used a blue LED with a peak wavelength of $\sim$470 nm. 
To identify the SEP one needs to gradually reduce the LED driving 
current until one reaches the state when the position of the spectrum does not 
depend on light intensity anymore.
Further reduction of the light intensity will then only change the proportion 
between the pedestal (which corresponds to events that do not release a
photoelectron from the photocathode) and the SEP.
With the SEP position known, the PMT gain at a given HV can be defined from:
\begin{equation}
\label{eq:sep}
{(N_{SEP}-N_{PED})\times q_0 = q_e \times G \times k },
\end{equation}
where $N_{SEP}$ and $N_{PED}$ denote the mean values of the signal and pedestal 
amplitudes in ADC channels,
$q_0$ is the charge per ADC channel~\footnote{e.g., 100 fC/channel for 
the CAEN V792 QDC}, $q_e\approx 1.602\times 10^{-19}~C$ is the electron 
charge, $G$ is the PMT gain at given HV, and $k$ is the signal attenuation 
or amplification factor between the PMT output and ADC input. To eliminate 
contribution from stray light, a ``Dark box'' was used for the PMT studies.
Fig.~\ref{xp4500-gain} illustrates the typical gain dependence on the high 
voltage for three randomly selected Photonis XP4500 PMTs.
The gain, on a logarithmic scale, depends linearly on the applied high voltage 
and varies from PMT to PMT.

%
In general, the precision of the gain measurements depends on the accuracy in 
the positions of the SEP and pedestal, as well as the distance between these 
two.  The uncertainty in these quantities in
our measurements is on average $\sim$10\%.

Once the gain of a PMT has been measured the relative quantum efficiency from 
a signal with a sufficiently large number of photons, {\it e.g.}, 
more than a few 100 photons, can be derived from the total charge 
collected from the PMT anode using
\begin{equation}
\label{eq:qe1}
{Q = N_{ADC} \times q_0 = N_{\gamma} \times QE \times G \times q_e },
\end{equation}
where $Q$ is the total charge, ${N_{ADC} = N_{SEP} - N_{PED}}$ is the peak 
value of the pedestal subtracted signal amplitude in ADC channels, 
${QE}$ and $N_{\gamma}$ denote the quantum efficiency and number of photons, 
and the other parameters are as in eq.~\ref{eq:sep}.

There are two ways to determine the quantum efficiency (QE) assuming a constant 
flux of incident light throughout the measurements. In the first method one
needs to know the PMT gain and can obtain the relative quantum
efficiency from eq.~\ref{eq:qe1},
\begin{equation}
\label{eq:qe2}
{ QE \sim N_{ADC}/G }.
\end{equation}
In the second method, the relative $QE$ of a PMT is obtained from the 
number of detected photoelectrons.
At fixed gain and fixed light intensity the number of detected photo-electrons 
depends only on the PMT quantum efficiency  $QE \propto N_{pe}$. 
Neglecting contributions from electronic noise and other possible fluctuations 
$N_{pe}$ can be estimated as inverse square of the 
normalized width of the detected photoelectron distribution,
\begin{equation}
\label{eq:sep1}
{ N_{pe} = 1/{\sigma_{norm}^2} },
\end{equation}
where
\begin{equation}
\label{eq:sep2}
 \sigma_{norm} = \sigma /N_{ADC}.
\end{equation}
with $\sigma$ the width of the amplitude distribution determined from a 
Gaussian fit and $N_{ADC}$ as before the pedestal-subtracted signal amplitude
in ADC channels.

For the relative quantum efficiency measurements the LED light intensity was 
fixed by setting the driving voltage. 
The PMTs HV were set to a fixed gain of $\approx2\times10^7$ (but different 
high voltage).

The number of photoelectrons detected can be obtained from both methods.
In the first, we first determined the SEP at a minimum LED 
intensity (at driving voltage $\sim$1 V) for each PMT. 
The PMT amplitude spectrum was then measured at high LED intensity (with a 
$\sim$4 V driving voltage, which corresponds to $\sim$6000 photons). 
In the second method, the LED intensity was kept fixed at 4.0 V and  the number
 of photoelectrons was determined from eq.~\ref{eq:sep1} and eq.\ref{eq:sep2}.
Before starting the relative quantum efficiency measurements, several systematic
tests to check linearity of the electronics,  LED stability, accuracy of the 
positioning of the PMTs relative to the LED, and reproducibility of the results
 were performed. The systematic uncertainty in the measurements due to these 
factors is about 5\%.
For each PMT both methods give a similar (within 15\%) number of detected
photoelectrons. 

At high LED intensities the XP4500 phototubes produce an 
average of $\sim$800 photoelectrons with a standard deviation of $\sim$25. 
This corresponds to a quantum efficiency of $QE \sim13-14\%$. 
For the XP4572 the average number of photoelectrons is $\sim$1100 with a 
standard deviation of $\sim$35, which corresponds to $QE \sim 18-20\%$.

\begin{figure}
\begin{center}
\epsfxsize=3.40in
\epsfysize=3.40in
\epsffile{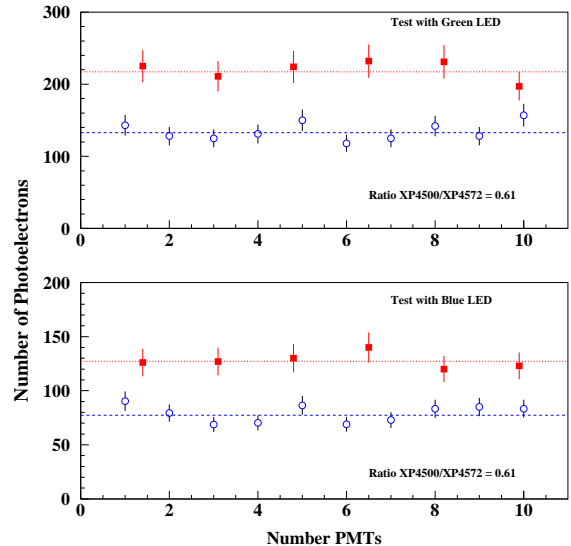}
\vspace{-0.5in}
\caption{\label{xp4500-xp4572-qe} (Color online) 
Relative quantum efficiencies (number of photoelectrons) for the XP4500 (open 
circles, blue) and XP4572 type PMTs (solid squares, red). 
The top (bottom) panel shows data measured with a green (blue) LED,
corresponding to intensities of  $\sim$1000 ($\sim$650) photons.
The XP4500 PMTs have on average $\sim$30-40\% less output signal as
compared to the XP4572 PMTs. }
\end{center}
\end{figure}

Figure~\ref{xp4500-xp4572-qe} shows the results from measurements of a randomly
selected sample of ten XP4500 and six XP4572 PMTs with both blue and green LEDs
 at light intensities of $\sim$1000 ($\sim$650) photons.
The XP4572 PMTs detected $\sim$220 ($\sim$130) photoelectrons, while the XP4500
 PMTs detected only $\sim$135 ($\sim$75) suggesting nearly 30-40\% less output 
signal.
Both types of PMTs have similar photocathode spectral sensitivity, at least in 
the region of blue and green light used in these measurements. 
The differences in the detected number of photoelectrons thus reflect a
real difference in quantum efficiency between these two PMT models.
Scans of the PMT photocathode show that the XP4572 effective 
area is much larger than that for XP4500 (see subsection~\ref{pmt-uniformity}), 
which may explain the observed difference.

The selection criteria for PMTs for the present detector are a combination of 
the i) highest gain at relatively low high voltage, and ii) the highest 
relative quantum efficiency. Thus, although the measured relative quantum 
efficiency for the XP4500 is lower than that of XP4572 PMTs their higher gain 
could, in principle, still make them suitable for the detector filled with 
aerogel $n$=1.030. 
However, for the lower refractive indices the light collection efficiency of 
the XP4500 is not suitable.

To check for possible improvements of the XP4500 light detection efficiency,
we explored the blue wavelength shifting (WLS) paint EJ-298\#2 for the PMT 
photocathode\footnote{The EJ-298\#2 absorbs light with 
wavelength ($\lambda$) in the region 300-400 nm and emits at 
$\lambda\sim$400-550 nm (with an efficiency of 70-90\%)~\cite{ElJen}.}.  
A thin ($\sim 100~\mu$m) and uniform layer of EJ-298\#2 was applied to 
the phototube window at room temperature.
Tests with such WLS for JLab Hall A gas \v{C}erenkov detectors have
shown improvements of $\sim$20\% for UV/Blue light~\cite{Allada}.

To test the effectiveness of the WLS the signal of one XP4500 PMT was 
measured with and without WLS. These measurements were performed using the 
single counter discussed in section~\ref{aero_study} with about ten tiles of
aerogel of thickness $\sim$10 cm and nominal refractive index $n$=1.030. 
No significant difference was observed in the output signal. 
This result may not be unexpected, since the aerogel material and the 
reflector on the walls of the single counter absorb more than 90\% of the 
produced UV light before it can reach the PMT.


\subsection{PMT photocathode uniformity}
\label{pmt-uniformity}

Uniformity of the signal across the surface of the photocathode is of concern 
for large diameter PMTs. A setup that can repeatedly reposition a directed 
light source, e.g., an LED, to different positions in front of the PMT was 
thus constructed to test the detection efficiency over the photocathode surface.

\begin{figure}[H]
\centering
{\includegraphics[height=2.5in]{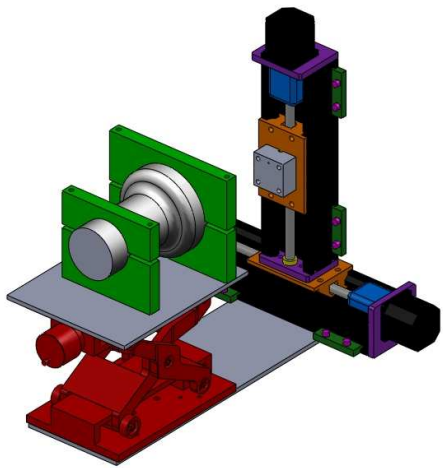} }
\caption{\label{pmt-cathod-scan} (Color online) Schematic of the PMT 
photocathode scanning system. The Velmex stepper motor setup allows for moving 
a blue LED along the two axes covering the full area of the photocathode. }
\end{figure}

The apparatus used for these measurements is shown in 
Fig.~\ref{pmt-cathod-scan}. A two-axis stepper motor setup (Velmex) was used 
which automatically repositions a blue LED attached to a 0.029-inch collimator 
and an optical fiber to direct the light to specific points on the PMT  
photocathode. 

An adjustable jack was used to position the PMT in the center of the rail's 
limits and the entire setup was installed inside a custom dark box.  
The motor controller was programmed to scan in 100 by 100 steps. 
At each point, the motor controller output triggers the DAQ system and a 
controllable pulse to the LED, collecting 30 data points per location. 
At each flash of the LED, an analog to digital converter integrates the current
 received by the PMT with a conversion factor of 0.25 pC. 
The stepper motor moved  450 steps at a step length of 0.0025 mm with every 
motor turn, resulting in a scan resolution of 1.2 mm. 
The total distance that each axis covered was 112.5 mm so that an effective
area of about $120~mm \times 120~mm$ of the PMT was scanned. 
The collected data were converted from the DAQ intrinsic CODA format to a plain
 ASCII format for further analysis.

\begin{figure}[H]
\centering
\subfigure[\label{pmt-4500-2d} PMT XP4500] 
{\includegraphics[width=3.4in,height=3.4in]{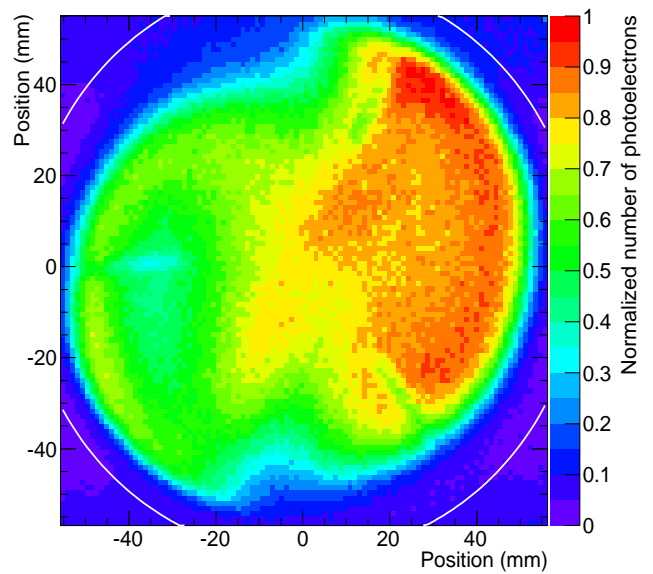} }
\subfigure[\label{pmt-4572-2d} PMT XP4572]
{\includegraphics[width=3.4in,height=3.4in]{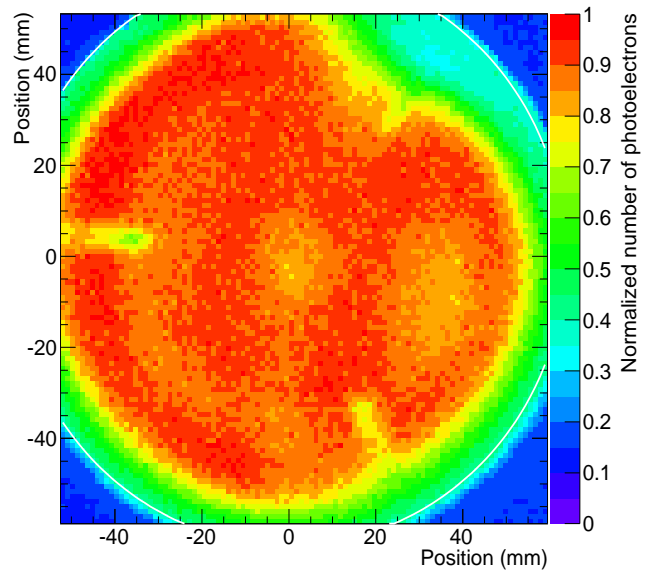} }
\caption{\label{pmt-2d-plots} (Color online) Two-dimensional scans 
of the number of detected photoelectrons (normalized) over the areas of
the XP4500 (top, a) and XP4572 (bottom, b) PMT photocathodes. }
\end{figure}

Fig~\ref{pmt-2d-plots} shows a typical two-dimensional effective photocathode 
area of the XP4500  (top panel, a) and XP4572 PMTs (bottom panel, b). 
For the XP4572 PMT, the region of highest efficiency covers almost the 
entire scanned region. In contrast, the region of highest efficiency for the 
XP4500 only covers about half the scanned area. The maximum difference of
the signals mean values across the photocathode is about 30\%.
This may explain our observation of low quantum efficiency for the XP4500 PMTs 
discussed in subsection \ref{pmt_gain_qe}.

In general, such non-uniformities of the photocathode response are not important for 
diffusive light collection as photons originating from the aerogel radiator 
reach the PMT surface only after several reflections and nearly uniformly 
illuminate the photocathode. Nevertheless, the non-uniformities need to be 
taken into account in the analysis, since it affects the PMT average quantum 
efficiency, and so the overall number of detected photoelectrons. 

\subsection{Magnetic shielding}
\label{magnetic-shield}

Photomultiplier tubes do not operate properly when exposed to magnetic fields. 
Magnetic fields may deflect electrons from their normal trajectory and cause a 
loss of photoelectron collection efficiency and gain.
The impact on the performance depends both on the PMT type, {\it e.g.}, their 
diameter and internal structure, and on the PMT orientation in the field. 
Covering the PMT with a cylindrical shield case longer than its length by at 
least half the shield case inner diameter is thus common practice.
The performance of this type of shielding in small magnetic fields, like the 
Earth's field, and a supplementary method to optimize the PMT performance
in larger magnetic fields is discussed below.

\subsubsection{Performance in weak external magnetic fields}
\label{earth-magnetic-shield}

For relatively small fields like the Earth field $\sim$0.3-0.6 G (30-60 
$\mu$T) $\mu$-metal typically provides sufficient shielding.
To study the effect of the Earth's magnetic field on the 5-inch PMT Photonis  
PMTs used in the SHMS aerogel \v{C}erenkov detector the PMT axis was 
oriented along the direction of the Earth magnetic field (axial). 
An LED-fiber coupled light source was used to illuminate the photocathode.
Data were taken both with the blue LED directly fixed on the photocathode, and
with the PMT at a distance of 10 cm from the light source to cover a wide 
surface of the photocathode as uniformly as possible. 
The LED intensity was kept constant during the measurement.

\begin{figure}
\begin{center}
\epsfxsize=3.50in
\epsfysize=2.50in
\epsffile{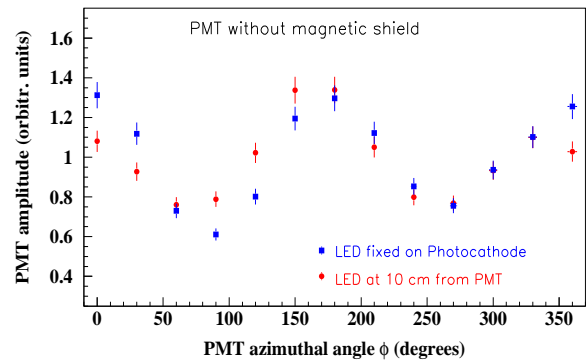}
\vspace{-0.5in}
\caption{\label{field-noshield} (Color online)
The PMT amplitude as a function of its orientation (measured in terms of the 
azimuthal angle $\phi$) in the Earth's magnetic field without magnetic 
shielding. Zero degrees is defined as the PMT axis oriented along the Earth
magnetic field. }
\end{center}
\end{figure}

In the first measurement the PMT without magnetic shield was rotated around its
axis (which is perpendicular to the photocathode) in $30^o$  steps clockwise  
to establish a baseline of the azimuthal dependence of its amplitude on the 
external field. The result is shown in Fig.~\ref{field-noshield}. 
The PMT amplitude depends strongly on the azimuthal angle varying in a 
$cos(2\phi$) pattern independent of the position of the light source, be it 
directly fixed on the photocathode or at a distance of 10 cm away from it.
Here, zero degree corresponds to the PMT pointing North. The azimuthal
variation corresponds to an amplitude roughly 40\% of the signal.

\begin{figure}
\begin{center}
\epsfxsize=3.50in
\epsfysize=2.50in
\epsffile{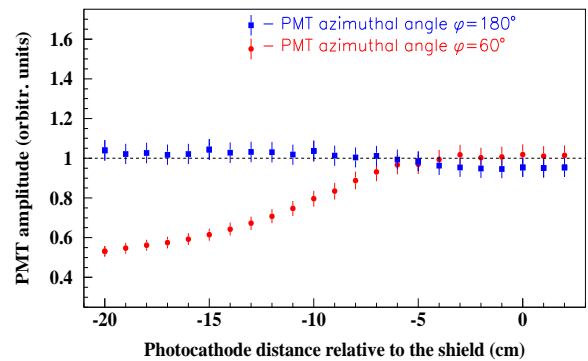}
\vspace{-0.5in}
\caption{\label{field-shield} (Color online)
The PMT amplitude as a function of photocathode distance relative to the shield
for its two orientation (azimuthal angle $\phi$=60 and 180 degrees) relative 
to Earth's magnetic field. }
\end{center}
\end{figure}

 For the subsequent measurements, the PMT was enclosed in a cylindrical 
magnetic $\mu$-metal shield of inner diameter $\sim$133 mm (5.236'') and 
thickness $\sim$1 mm (0.040''). In these measurements, the degree to which the 
PMT could still be affected by the external field was tested by shifting the 
photocathode relative to the edge of the magnetic shielding cylinder. 
A distance of "zero" means that the PMT photocathode is positioned right at the
 edge of the magnetic shielding cylinder. The horizontal axis is perpendicular 
to the PMT photocathode surface and measurements were performed for two 
orientations of the  PMT with respect to the Earth's field, at azimuthal 
angles of 180 and 60 degrees, respectively. 
Here, 180 degrees corresponds to an orientation parallel, but in opposite 
direction to the Earth's magnetic field, while 60 degrees is a large angle near
 perpendicular. The performance of the magnetic shield for all cases is 
illustrated in Fig.~\ref{field-shield}. 

For the photocathode located outside of the magnetic shielding up to distances 
of $\sim$5 cm, the Earth's field does not have any notable impact on the PMT. 
For the photocathode located fully within the magnetic shield cylinder (zero or
 positive distances), or even moving the magnetic shield cylinder a distance of
 5 cm behind the photocathode, the weak magnetic field has no effect, for 
either angular orientation, and the magnetic shielding remains effective. 
For an orientation of $\phi$=180 degrees the magnetic shielding remains even up
 to a distance of the magnetic shield cylinder 20 cm behind the photocathode, 
there is no effect independent of the photocathode's position inside the 
shielding cylinder. On the other hand, moving the photocathode more than 5 cm 
outside of the magnetic shielding for an azimuthal angle of $\phi$=60 degrees
has a large effect: the effect of the Earth's magnetic field can again reduce 
the amplitude, by up to 40\%, to closely mimic the unshielded situation.

These results have an important design impact for our detector, the magnetic 
shield of the PMTs only need to be extended up to edge of the photocathode. 
Based on the results of the studies the choice of a $\sim$1 mm (0.040'') 
thickness $\mu$-metal  cylinder with inner diameter of $\sim$133 mm (5.236'') 
is confirmed, in order to shield the PMTs of the present detector from weak 
external magnetic fields ($<$ 1 Gauss). The PMT magnetic shielding is aligned 
with the edge of the photocathode. 

\subsubsection{Performance in stronger external magnetic fields}
\label{bucking-coil}

In the SHMS detector enclosure where the aerogel \v{C}erenkov detector will be 
located fringe fields from the spectrometer magnets are expected to be less 
than 0.5 Gauss.  
However, the design of the present detector is flexible and includes a method 
to allow PMT operation in fields up to 5 Gauss. This would allow operation of 
the detector at a location in the detector stack that is close to the SHMS 
dipole, {\it e.g.}, before the first hodoscope plane to further reduce knock-on
 or $\delta$-electrons, should it be needed for future experiments.

An effective technique to minimize the effect of spectrometer residual magnetic
fields is the use of bucking coil systems~\cite{Gogami}. The current sent 
through the bucking coil produces a magnetic field opposite in direction to the
external magnetic field from the spectrometer. By regulating the coil current 
one can thus reduce (compensate) the effect of external fields on the PMTs.

\begin{figure}[H]
\begin{center}
\epsfxsize=3.10in
\epsfysize=2.60in
\epsffile{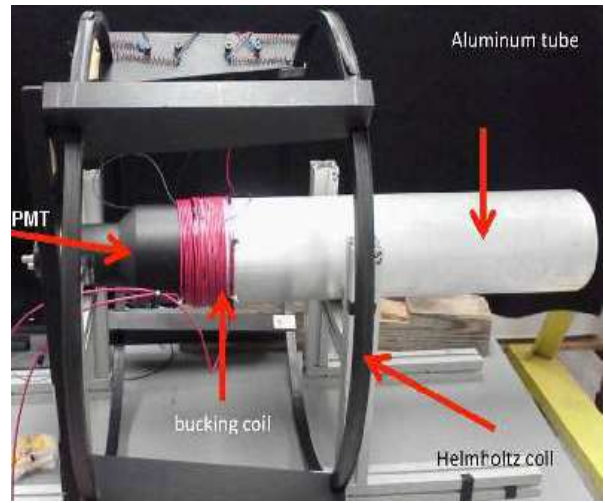}
\caption{\label{bucking-coil-setup} (Color online)
Experimental setup developed for bucking coils studies.}
\end{center}
\end{figure}

The experimental setup used for the bucking coil studies is shown in 
Fig.~\ref{bucking-coil-setup}.
The PMT and its magnetic shield were placed inside an aluminum tube of the same
 type as those that are used to hold the PMTs in place in the aerogel 
\v{C}erenkov detector itself. 
The external magnetic field simulating the spectrometer residual fields was 
provided by a Helmholtz coil producing a magnetic field of about 5 Gauss.
In this setup the magnetic field is in linear relationship with the current 
through the Helmholtz coil., which allows for calculating the field for a given 
current directly. 

Without external magnetic field the PMT amplitude depends logarithmically on 
the operating high voltage. When the magnetic field was applied, the PMT gain 
(driven by the secondary emission of the dynodes) was not significantly 
affected for field strengths ranging between 1-10 Gauss.  The photoelectron 
collection efficiency (from photocathode to first dynode), however, decreases 
significantly even for relatively small fields of 1-2 Gauss resulting in a 
20-50\% loss of signal. This confirms that the decrease in PMT efficiency is 
due to photoelectrons acted on by the magnetic field before they arrive at the 
dynode cascade. 

\begin{figure}[H]
\begin{center}
\epsfxsize=3.4in
\epsfysize=3.0in
\epsffile{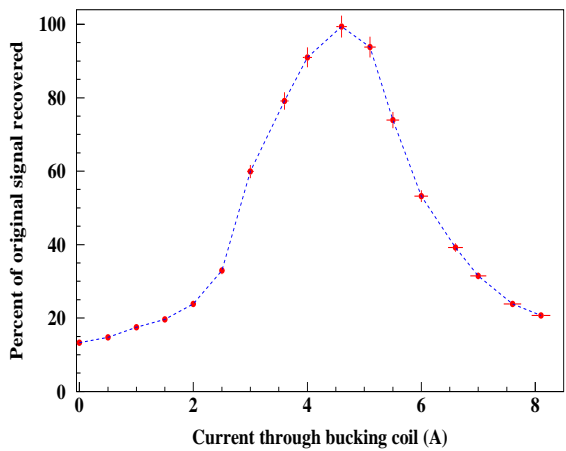}
\vspace{-0.5in}
\caption{\label{bucking-coil-results} (Color online)
Recovery of the PMT signal in an external (few-Gauss) field by applying current
 to a bucking coil. When the external and bucking coil fields cancel, the 
signal is fully recovered.}
\end{center}
\end{figure}

The PMT signal recovery using a bucking coil system is illustrated in 
Fig.~\ref{bucking-coil-results}. For the signal to be fully recovered, the net 
magnetic field inside the PMT needs to be zero. This happens in our setup 
roughly when a current of 4.6 A is applied to the 20-turn bucking coil, allowing
 for an almost 100\% signal recovery.
Up to an applied current to the bucking coil of about 2 A the net magnetic 
field is large due to the external 5 G magnetic field applied, and most of the 
signal is lost. As the bucking coil current increases beyond 2 A, the magnetic 
field it produces eliminates part of the external field, until their magnitudes
 are matched and they cancel each other achieving a net field of zero at the 
$\sim$4.6 A current. 
Once the field produced by the bucking coil exceeds that of the external 
magnetic field, at bucking coil current above 4.6 A, the signal recovery starts
 to decrease again as the net field inside the PMT becomes larger. 
For the present detector the setup thus includes a flexible design allowing to 
add the bucking coil feature, to fully compensate for external fields
 of about 5-10 G.

\section{Detector Performance}
\label{perform}

The efficiency and background rejection capability of the aerogel \v{C}erenkov 
detector depends on the geometry optimization, on the thickness of the 
radiator, and the location in the SHMS detector stack.
Extensive Monte Carlo studies and tests with a prototype of the SHMS aerogel 
\v{C}erenkov detector were performed to optimize  the aerogel thickness for the
 best kaon detection efficiency and highest rejection factor for protons. 
Details of the Monte Carlo (MC) program and the material parameters used can be
 found in section~\ref{mc_sim}. 
For quality assurance and to verify light collection and uniformity, the full 
detector was tested with cosmic rays before and during the filling of the 
aerogel trays, and again after the installation in the SHMS detector hut. 
This section describes the detector performance including detection efficiency 
and background rejection, sub-threshold background rejection, as well as light 
collection and uniformity.

\subsection{Projected kaon detection efficiency and proton suppression}
\label{aero-projection}

\begin{figure}[H]
\begin{center}
\epsfxsize=3.40in
\epsfysize=3.00in
\epsffile{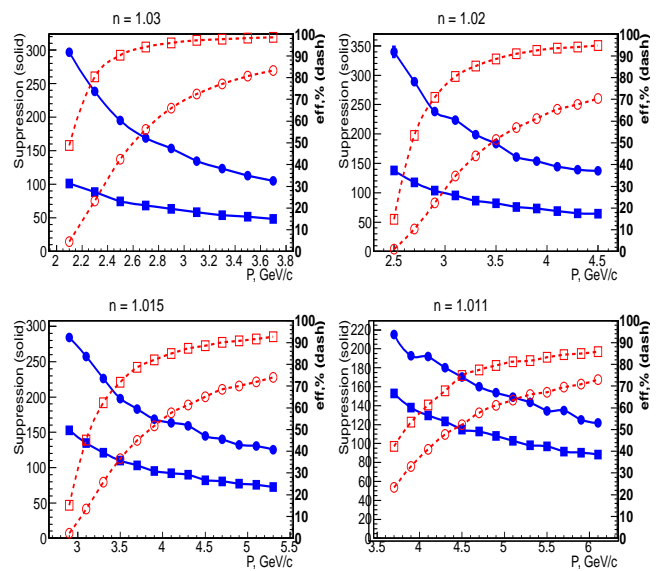}
\caption{\label{fig:suppression} (Color online) 
 The projected kaon detection efficiency (open symbols) and proton 
suppression (solid symbols) versus particle momentum for the SHMS Aerogel 
\v{C}erenkov Detector with nominal refractive indices of n=1.030,~1,020,~1.015 
and 1.011. 
The open (solid) squares represent projections for a cut on the number of 
photoelectrons of $N_{pe}>$1. 
The open (solid) circles represent the projections for cuts $N_{pe}>$5 (SP-30), 
$N_{pe}>$4 (SP-20), $N_{pe}>$3 (SP-15), and $N_{pe}>$2 (SP-11). }
\end{center}
\end{figure}

The ultimate performance of the detector for kaon detection efficiency is 
correlated with its performance for proton suppression. One can reach higher 
kaon detection efficiency at the cost of less proton suppression. This choice 
will be based on the actual kaon to proton ratios as present in the experimental 
configuration. 
Some examples are presented in Fig.~\ref{fig:suppression}, which illustrates 
the projected kaon detection efficiency and proton suppression for the SHMS 
Aerogel \v{C}erenkov Detector as a function of particle momentum for two 
different cuts on the number of photoelectrons. 
The four panels represent projections for the different refractive indices, 
$n$ = 1.030, 1.020, 1.015 and 1.011, respectively.

For the regions of interest defined by the threshold momenta listed in Table I,
the kaon detection efficiency for a cut $N_{pe}>$1 (open squares) is no less 
than 90\% for all refractive indices except for the lowest refractive index, 
where it is about 85\%.  The proton suppression (open squares) is on the order 
of 50-70:1 in the respective region of interest for all refractive indices 
except for the SP-11. There, the projected proton suppression is about 100:1 
for momenta above 5.4 GeV/$c$ 
The projected detection efficiency and proton suppression are suitable for the 
experiments described in section~\ref{intro}.

The kaon signal detection efficiency and proton rejection factor can be further
optimized by requiring different (stricter) cuts on the number of  
photoelectrons. 
As an example the open (solid) circles represent the projections for cuts on
$N_{pe}>$5 (SP-30), $N_{pe}>$4 (SP-20), $N_{pe}>$3 (SP-15), and 
$N_{pe}>$2 (SP-11). 
For these cuts, the proton rejection factor increases to about 150:1. However, 
this comes at the expense of a decreased kaon signal detection efficiency, 
which would be reduced to 70\% on average in the region of interest. 

\subsection{Sub-threshold background rejection}
\label{knock-e}

The main limitation for the background rejection capability of threshold 
detectors is the production of knock-on electrons (also called 
{\it $\delta$-electrons}).
The general impact of $\delta$-electrons on aerogel detector rejection 
capabilities is discussed in Refs.~\cite{Nishida,Coman}. 
There, the impact of $\delta$-electrons on a measurement was estimated to be a 
few percent depending on geometry, details of the construction and location of 
the detectors.
In general, the probability to produce $\delta$-electrons by particles with a 
momentum below the \v{C}erenkov radiation threshold was shown to increase with 
radiator thickness.

The SHMS Aerogel \v{C}erenkov Detector is located after the noble-gas 
\v{C}erenkov, the pair of drift chambers, the first pair of scintillation 
hodoscope planes (1X and 1Y) and the heavy gas \v{C}erenkov detector. 
A non-negligible amount of $\delta$-electrons will be produced in the materials
 of these detectors and in the air-gaps between them, and some of these may 
have enough energy to penetrate inside the effective volume of the aerogel 
detector and produce themselves \v{C}erenkov light.
Thus, protons with a momentum below the threshold momentum for generating 
\v{C}erenkov light in aerogel may produce $\delta$-electrons in the detectors 
located upstream of the aerogel \v{C}erenkov counter or/and in first several 
layers of the aerogel itself.  
Since a diffusion-type aerogel detector only provides information about the 
number of detected photons, there is no way to separate these events from the 
real kaon events. 
These $\delta$-electrons may reduce the rejection capability of the aerogel 
detector and contribute  to fluctuations in the signal (see for example 
Ref.~\cite{Grove}), and thus merit a more detailed discussion. 

We have estimated the number of $\delta$-electrons which can be generated in 
the SHMS Aerogel \v{C}erenkov Detector. 
For simplicity, first we took into account the aerogel detector alone. 
For a conservative estimate we assumed an entrance window thickness of 1.5 mm 
(about 30\%  thicker than the actual thickness of the detector entrance window).

In general, the spectrum of $\delta$ electrons that have sufficiently high 
energy to produce \v{C}erenkov light is given by~\cite{Seltzer-Berger}
\begin{equation}
\label{eq:knock}
N(E)dE = 0.30058 \frac{mc^2}{\beta^2} \frac{Z}{A}  
(1-\frac{\beta^2 E} {E_m}) \frac{1}{E^2} dE,
\end{equation}
where $N(E)$ is the number of $\delta$-electrons of kinetic energy $E$ produced 
per $g/cm^2$ of a target of given ($Z$/$A$), $\beta$ is the proton velocity, 
$m$ is the electron mass, and $E_m$ is the maximum kinetic energy transferred to
the $\delta$-electron in an ion-electron collision.

The $\delta$-electron spectrum is falling rapidly at electron energies near the
 \v{C}erenkov threshold. The total number of $\delta$-electrons capable of 
producing \v{C}erenkov light increases slowly with the primary energy. 
However, the total energy contained in them rises rapidly as does the average 
energy~\cite{Evenson}. The total light yield generated by these 
$\delta$-electrons can be calculated, in principle, taking into account
 the light yield of stopped electrons, which is for most materials near minimum
 ionizing until it is below threshold. 
However, in practice this becomes complicated. Significant contributions from 
$\delta$-electrons to the signal can originate from electrons with energies 
well beyond threshold, of 20 MeV or more~\cite{Evenson}. To include all 
cascading effects in all materials for such higher-energy electrons would 
require a solution to the transport equation, which is beyond the scope of this
 article. 

\begin{figure}[H]
\centering
\includegraphics[width=3.5in,height=2.5in]{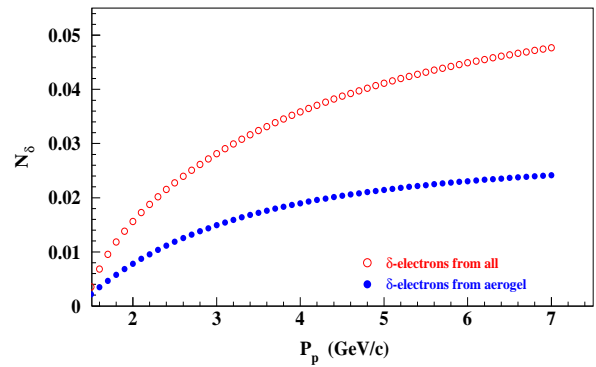}
\vspace{-0.5in}
\caption{\label{knock-aero} (Color online) Probability to produce 
$\delta$-electrons versus proton momentum. The solid (blue) circles are for a
standalone aerogel detector. The open (red) circles take into account all 
materials of windows and detectors in the shielded SHMS detector hut before the
 location of the aerogel detector.}
\end{figure}

Instead, we make the approximation that below some cut-off energy the 
$\delta$-electrons contribute according to equation~\ref{eq:knock} while above 
this energy the contribution equals that of an electron with the cut-off 
energy. We require the $\delta$-electrons produced to be within the angular 
acceptance defined by both the distance to the detector and the aerogel 
detector effective area, and applied a $E >2$ MeV cut of energy. 
The threshold momentum for electrons in aerogel of nominal refractive index 
$n$=1.030 is 1.62 MeV, so with a 2 MeV energy cut-off we require this to be 
slightly above the threshold. The results are shown in Fig.~\ref{knock-aero}.

In general, the number of $\delta$ electrons increases with material thickness.
In  Fig.~\ref{knock-aero} we show both the probability for protons in the 
2-7 GeV/$c$ momentum range to produce $\delta$-electrons in the aerogel 
detector as a standalone detector (closed circles) and taking into account all 
materials in the shielded SHMS detector hut before the aerogel detector 
location (open circles). The probability grows with increasing momentum as 
expected.

The probabilities shown are most likely an overestimate. 
To estimate the actual effect of $\delta$-electrons on the experiments one has 
to also take into account the particle identification properties of the other 
detectors available in the SHMS. In particular, the SHMS includes a segmented 
lead glass calorimeter consisting of a preshower of 3.6 radiation lengths and a
 shower counter of 18.2 radiation lengths located behind the aerogel detector. 
Information from these can also be used to estimate or constrain the effect of 
$\delta$-electrons on the experiment. 
For example, the hadron/electron rejection of the shower and preshower counter 
$\sim$250:1 can be achieved in the momentum range of 2 to 11 GeV/$c$ for 
$\sim95-99\%$ electron detection efficiency~\cite{Mkrtchyan}.
The segmentation of the calorimeter also allows for taking advantage of the 
differences in the spatial development of the electromagnetic and hadronic 
showers. 
The electromagnetic showers develop generally earlier and deposit more energy 
at the beginning than hadrons. Thus, a measurement of the energy deposited at 
the front of the calorimeter along with the total energy deposition can further
 improve the electron/hadron separation.

\subsection{Light Collection Performance and Uniformity}
\label{cosmic_tests}

\subsubsection{Optimization of the light collection}
\label{light_collection_efficiency}

Good light collection efficiency is an important aspect for the SHMS aerogel 
\v{C}erenkov detector performance. This is particularly important for the light
 collection efficiency of the detector's lowest nominal refractive index 
(SP-11), where the expected signal is small, as shown from the calculations in ~\ref{subsec:aero_LightYield}.

\begin{figure}[H]
\centering
\includegraphics[width=3.5in]{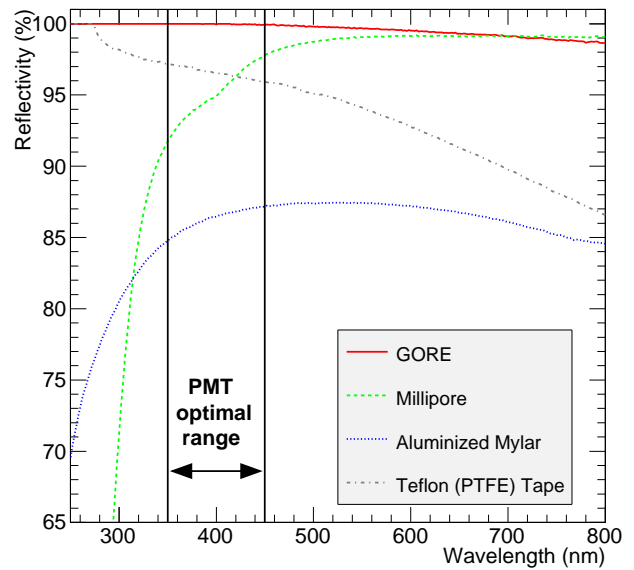}
\caption{\label{fig:reflectors} (Color online) Reflectivity of several materials commonly used for covering the interior of detectors. Gore diffusive reflector material has the highest reflectivity where PMTs are most sensitive ($\lambda\sim$350-450 nm) followed by Millipore. In the SHMS Aerogel detector, we used Millipore to cover the trays with aerogel refractive index $n$=1.030 and $n$=1.020, and GORE reflector for the trays with $n$=1.015 and $n$=1.011 to optimize the light collection.}
\end{figure}

One possible optimization is that of the reflector material in the diffusion box and in the trays. Millipore is the most economical and most commonly used reflector material, however its reflectivity is around 96\% in the wavelength range of the PMTs ($\lambda\sim$350-450 nm) and one expects several light bounces in the detector. 
We studied the reflectivity of several materials often used for detector construction, such as teflon, aluminized mylar, and GORE reflector, aiming to improve the light collection for the lowest index trays and the diffusion box. Measurements of reflectivity of these materials were done using the reflectance port of the integrating sphere shown in Fig.~\ref{fig:intSphere}. Figure~\ref{fig:reflectors} compares the measured reflectivity of GORE (1~mm) and Millipore, highlighting the spectral region of high efficiency of the PMTs. The 1 mm thick Gore diffusive reflector material (available in sheet sizes $12''\times12''$) has a reflectivity greater than 99.7\%. The also available 3.2 mm thick reflector material provides an even higher reflectance~\cite{Gore}.  

\begin{figure}[H]
\centering
\includegraphics[width=\columnwidth,height=3.0in]{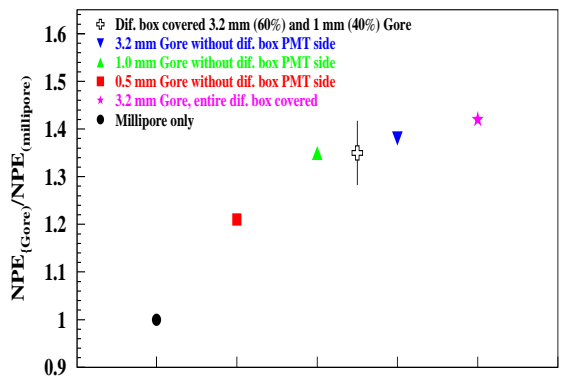}
\vspace{-0.5in}
\caption{\label{gore_mili} (Color online) Ratio of number of photoelectrons as 
predicted by  simulation for lining (sections of) the detector with different 
thickness of Gore reflective material relative to lining the detector with 
Millipore. The open cross shows an experimental measurement for the final 
chosen configuration with the detector covered for 60 (40)\% with 3.2 (1.0) mm 
thick Gore reflective material, in good agreement with the simulated ratios.}
\end{figure}

Fig.~\ref{gore_mili} shows a Monte Carlo simulation of the detector performance
 if one covers the detector with Gore reflector of different thicknesses, as a 
ratio to the base detector design with millipore. In some cases only sections 
of the detector was covered with the Gore reflective material to illustrate 
that the sensitivity to the higher reflective material is not uniform over the 
inner detector area. 
The simulation studies suggest that lining the sides of the diffusion box where
 the PMTs are housed with the higher reflective material does not make much
difference, and that much of the improvement in light collection efficiency is 
already reached by lining the detector with Gore reflective material of 1 mm 
thickness. Covering the surface of the backplane of the detector's diffusion 
box with 0.5 mm thick Gore reflective material already gives a little over 20\%
 improvement in light collection efficiency as compared to the base design with
 Millipore only. The ultimate performance is reached by covering the entire 
diffusion box with Gore reflective material of 3.2 mm thickness, rendering a 
little over 40\% improvement in light collection efficiency as compared to the 
base design. We chose an intermediate solution and lined 60\% of the diffusion 
box with 3.2 mm and 40\% with 1 mm thick Gore reflective material.

The average number of detected photoelectrons after (before) replacing 
Millipore with Gore diffusive reflector in the entire diffusion box is $\sim$12
 ($\sim$9) for the aerogel with refractive index $n$ = 1.030 (SP-30) and and 
$\sim$8.0 ($\sim$6) for $n$ = 1.020 (SP-20). 
The use of the XP4572 PMTs instead of XP4500 in the diffusion box in turn
improved the performance of the detector by $\sim$35--40\% 
(see section~\ref{pmt_gain_qe}). 
Thus, Gore reflective material and XP4572 phototube have generally resulted in 
improving the characteristics of the detector by $\sim$70\%.      
This improvement is most significant for the two lower refractive 
aerogel index trays,  where the expected number of photoelectons is small.

\subsubsection{Measurements with beam}
\label{beam_tests_aerogel}

For a confirmation of the light collection efficiency, we tested the aerogel 
with nominal refractive index of $n$=1.030, one of the aerogel materials 
previously used in experiments at MIT-Bates, during the P-349 experiment at 
CERN, whose first run was completed successfully in December 2014.
The particles of interest in this experiment are antiprotons of 
momentum 3.5 GeV/$c$ produced by the 24 GeV/$c$ momentum proton beam of the 
CERN/PS along with other secondary particles of the same momentum. 
These antiprotons have a velocity of $\beta$=0.966. Pions of the same momentum 
have a velocity close to the speed of light, $\beta$=0.9992.
The aerogel thus allows for discriminating these hadrons.
Initial analysis suggests the aerogel to indeed perform as expected lending 
further confidence for the foreseen Aerogel \v{C}erenkov detector's
performance in the SHMS. 

\subsubsection{Measurements with Cosmic Rays}
\label{cosmic_tests_detector}

For an initial check, both during the assembly of the various aerogel trays,
layer by layer, and after final assembly before installation, the SHMS Aerogel
\v{C}erenkov Detector was tested with cosmic rays.
For these tests, the detector was positioned horizontally, with the diffusion 
box on top of the aerogel tray. Two trigger scintillator counters
aligned with respect to each other were positioned above and below the detector.

The signal pulses from the scintillators were discriminated and fed into a 
Logic Unit to form a coincidence trigger providing a ''Gate'' for a CAEN V792
charge integrating QDC module, or alternatively as a ''Start'' for a TDC 
module.  
The QDC sensitivity was set to 100 fC/channel.  The signals generated by
cosmic ray muons from all 14 PMTs (7 from each side of the diffusion
box) were sent to a passive splitter (50:50). 
Then, one output was sent to the QDC module through a $\sim$150 ns delay line 
to fit within the gate generated by the trigger scintillators. 
The second output was sent to a discriminator module
(Phillips 708) and then to a TDC module. The latter can be used for timing 
measurements, but these were not the main goal of the current tests. 
All information from the QDC were read out by a CODA data acquisition system. 
The offline analysis was done using the ROOT analysis package.

We have checked the stability of the results - the total sum of photoelectrons,
with respect to (i) the ADC gate width by changing it from 150 ns to 200 ns;
(ii) the timing of the individual PMT signals relative to the gate by shifting 
signals by $\pm$50 ns; and (iii) the gain of PMTs by changing high voltages by
$\pm$100 V.  No significant variation in the total sum of photoelectrons
measured was found.

\begin{figure}[H]
\centering
\includegraphics[width=\columnwidth]{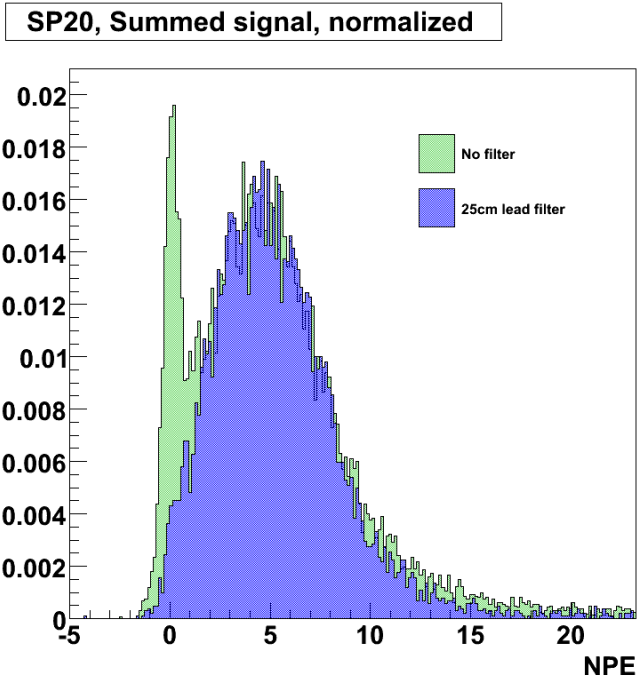}
\caption{\label{sp20_lead} (Color online) Signal distributions from
  cosmic rays passing through the center of the SHMS aerogel detector filled
  with 7 layers (7.7~cm) of SP-20 type aerogel, both with (blue) and without 
(green) a 25~cm lead stack acting as low-energy filter placed between the 
trigger scintillators. }
\end{figure}

The effect of the low-energy components of cosmic rays on the data were first 
checked, by comparing measurements with and without a 25 cm thick lead filter 
installed above the lower trigger scintillator. Our measurements show that the 
total signal amplitude did not change significantly after filtering out these 
low-energy components (see Fig.~\ref{sp20_lead}).

\begin{figure}[H]
\centering
\includegraphics[width=\columnwidth]{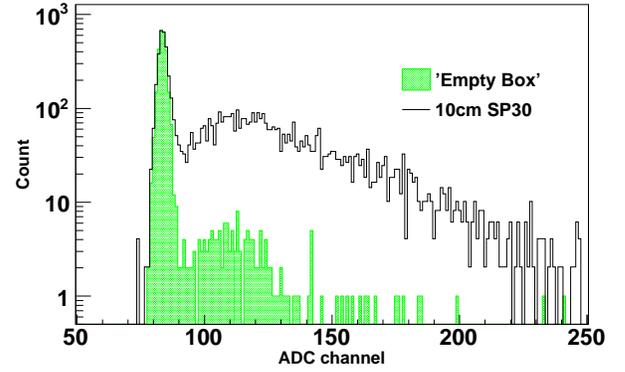}
\caption{\label{adc_spec2} (Color online) Representative ADC spectra 
of a detector PMT, triggered by cosmic rays passing through an area where
a 10 cm thickness of SP-30, $n$=1.030, aerogel material was stacked and
an area where no aerogel radiator material was stacked ('Empty Box').
The histograms are normalized to a common maximum height.}
\end{figure}

The level of background was determined using the cosmic rays passing through 
empty space - an area where there were no aerogel tiles stacked - in the 
detector. Along with pedestal events, the single photoelectron signal was 
detected, albeit with low intensity, in all the detector PMTs indicating 
scintillation in air and reflector materials within the detector.  
These signals became more prominent when \v{C}erenkov light from the cosmic 
rays passed through a 10 cm thick aerogel radiator of refractive
index $n$=1.030. Fig.~\ref{adc_spec2} shows a comparison of the two signals.

\begin{figure}[H]
\centering
\includegraphics[width=\columnwidth]{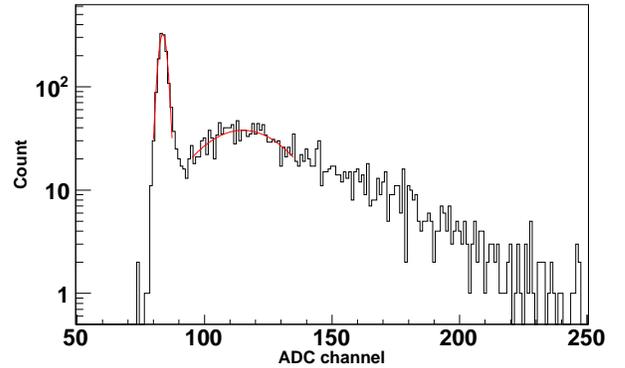}
\caption{\label{adc_spec} (Color online) Typical example of a Gaussian fit to
the ADC spectrum of one of the detector PMTs. The mean values of these
Gaussian fits allow one to identify the pedestal and single photoelectron
peak positions.}
\end{figure}

Fig.~\ref{adc_spec} illustrates a double Gaussian fit to a typical ADC spectrum
of the detector PMT signals, to identify both the pedestal and the single 
photoelectron (SPE) peak positions. The total (summed over all channels) 
detector signal was obtained from the pedestal subtracted SPE position.
This then allows us, after summing all detector PMTs, to convert the measured
ADC spectra to the anticipated number of photoelectrons used in further figures.

\begin{figure}[H]
\centering
\includegraphics[width=\columnwidth]{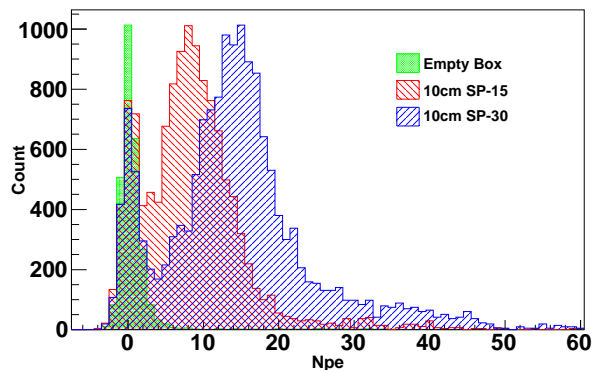}
\caption{\label{refind_test} (Color online) The summed detector PMT signals
  from cosmic rays passing through a 10 cm stack of SP-15 (red) and SP-30
  aerogel (blue) positioned at the center of the aerogel box. The
  green shaded area corresponds to the rays passing through an empty region
  outside the aerogel radiator, within the detector. 
The histograms are normalized to a common maximum height.}
\end{figure}

We first placed a single stack of aerogel tiles in the center of the aerogel 
box. Thus, the stack's area was that of a single tile, i.e. $11\times~11~cm^2$ .
The light collection-performance of the detector in this configuration was 
tested by taking data with aerogel tiles  stacked up to 10 cm high. 
Fig.~\ref{refind_test}, shows the results, converted to the number of 
photoelectrons, for two nominal refractive indices, $n$=1.015 and 1.030, in 
comparison to the case with no aerogel material present. The signals for the 
two refractive indices differ by a factor of $\sim$2, as  expected from 
analytical calculations. Cosmic rays passing through an area where no aerogel 
is stacked  ('empty box') produce a, slightly asymmetric, peak at zero with a 
width of $\sim$1 p.e. (HWHM).

\begin{figure}[H]
\centering
\includegraphics[width=\columnwidth]{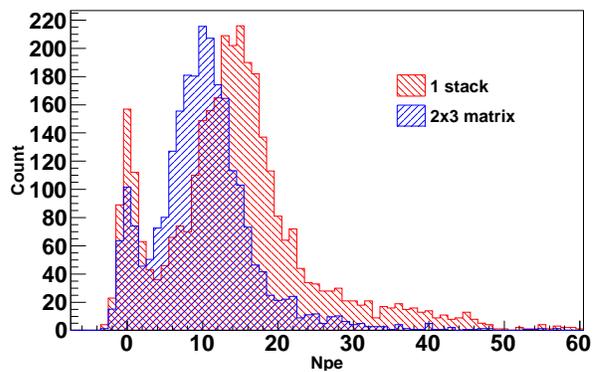}
\caption{\label{absorp_test} (Color online) Response of the detector
  to the cosmic rays passing through a 10 cm thick single stack of
  aerogel with refractive index $n$=1.030 at the center of the aerogel box 
  (red), and through a matrix of 2$\times$3 aerogel stacks (blue). 
  The histograms are normalized to a common maximum height.}
\end{figure}

The above case somewhat resembles the single counter prototype discussed in
section~\ref{aero_study}. The difference in light absorption of the full 
diffusion box as compared to the single counter prototype was tested by next
loading the detector box with different volumes of the aerogel material. 
Fig.~\ref{absorp_test} illustrates the detector response for a 10 cm thick 
single stack and a 2$\times$3 aerogel stack. The data suggest a signal reduction
of $\sim$30\% when the aerogel volume is increased 6-fold, in a matrix of two by
three vertical aerogel stacks.

Naively one may assume that increasing the aerogel material volume will always 
yield more photoelectrons, and thus increase the detection efficiency. 
However, increasing the aerogel also increases the scattering and attenuation 
by absorption of photons. 
This was already observed in section~\ref{subsec:aero_LightYield} for the case 
of the single counter prototype, where the signal was shown to saturate with 
increasing aerogel thickness. This is as expected, due to scattering and 
absorption of the light by aerogel at short wavelengths.

\begin{figure}[H]
\centering
\includegraphics[width=\columnwidth]{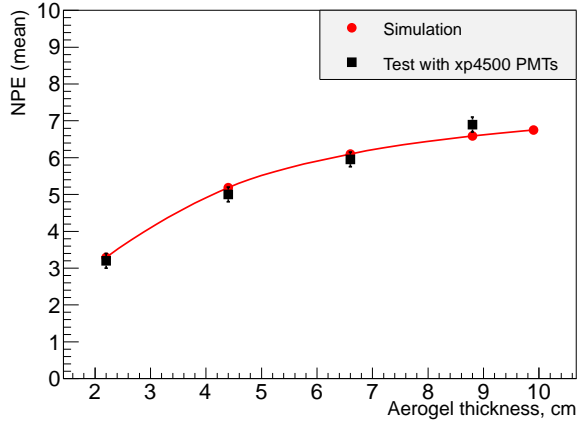}
\caption{\label{sp30_filling} (Color online) The summed detector PMT signals
  from cosmic rays passing through the center of the detector as a function of
  the aerogel thickness. The data were obtained while filling the aerogel tray 
with aerogel with refractive index $n$=1.030 (SP-30). The solid squares 
indicate the measurements. The solid circles, and the line, denote the 
projections from the Monte Carlo simulations and the best second-order 
polynomial fit, and are in excellent agreement with the measurements. 
Note that the number of photoelectrons increases to $\sim$11 (for an 
aerogel thickness of 8.8 cm) if using XP4572 PMTs and lining of the aerogel 
diffusion box with Gore reflective material, see text.}
\end{figure}

Thus, finding the optimal aerogel material thickness is a balance at which one 
tries to keep the effect of light attenuation at a minimum, 
maximize the radiation of light, and reduce the amount of aerogel to minimize
the production of $\delta$-electrons in it. 
After this threshold, 
the effects of attenuation begin to out-weight the gain of additional 
photoelectrons.
To test the performance of the aerogel radiator in the full \v{C}erenkov 
detector as a function of radiator thickness, we carried out measurements with 
cosmic rays after installing every other aerogel layer during the tray 
assembly. Fig.~\ref{sp30_filling} shows the results for aerogel with nominal 
refractive index $n$=1.030. Note that during these test measurements the 
aerogel detector was lined with Millipore reflector and was equipped with 
XP4500 PMTs.
In the simulation results, the PMT quantum efficiencies were also degraded by 
a factor of 0.61 in accordance with the findings in our PMT studies 
(see section~\ref{pmt_study}). The measurements are in agreement 
with the projections from the simulations.

At the optimal thickness of 8 layers (8.8 cm) of SP-30 aerogel, the summed 
signal from cosmic rays passing through the center of the detector has a peak 
value of $\sim$7 photoelectrons.
From here on, all further tests are performed with the aerogel \v{C}erenlov 
detector diffusion box equipped with XP4572 PMTs and lined with a combination 
of 3.2 mm thick and 1.0 mm thick Gore reflective material, as described above, 
which raises the observed photoelectrons in this case to $\sim$11 
photoelectrons.

Finally, we checked the spatial uniformity of the detector response by 
measuring the coordinate dependence of the detector signals, always with trays 
filled with 8 layers of aerogel (8.8 cm thickness of aerogel material). 
For these tests the trigger scintillators were moved along two median lines: 
a horizontal scan from the left middle PMT to the right middle PMT, and a 
vertical scan from top to bottom along the line in the middle of the left and 
right PMTs of the detector. 
For each scan, the $\sim$3'' ($\sim$8 cm) wide and $\sim$16'' ($\sim$40 cm) 
long trigger scintillators were oriented transversely to the scan directions 
in order to minimize uncertainty in position.

\begin{figure}[H]
\centering
\includegraphics[width=\columnwidth,height=2.5in]{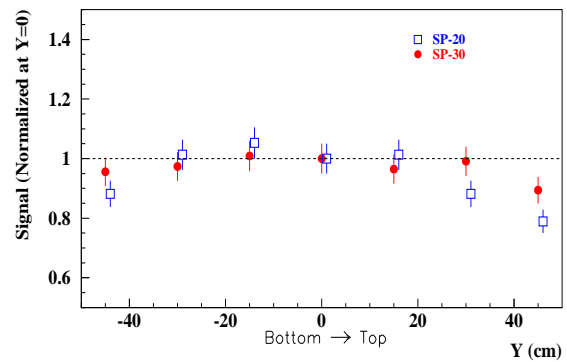}
\vspace{-0.5in}
\caption{\label{v_scan} (Color online) Dependence of the summed signal
  from the SHMS aerogel detector on the impact point position of
  cosmic rays on the top -- bottom median line of the
  detector. Signals from the detector with tray SP-30 and SP-20 are normalized 
  at Y=0.0 cm. }
\end{figure}

The vertical scans did not elucidate much: they do not show
significant coordinate dependence of the signal, similar to the earlier 
observations with the HMS aerogel detector~\cite{Asaturyan}. 
Figure~\ref{v_scan} shows as an example the results from the vertical scan of 
the SP-30 and SP-20 trays, with the signals normalized at the center (Y=0 cm).

\begin{figure}
\centering
\includegraphics[width=\columnwidth]{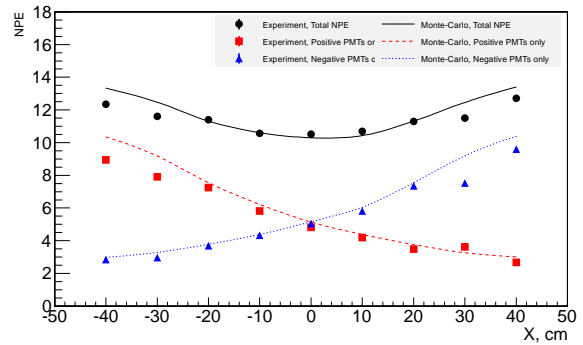}
\caption{\label{h_scan} (Color online) Dependence of signals from the
  SHMS aerogel detector on the impact point position of cosmic rays on
  the left -- right median line of the detector (see text). The tray is filled
  with 8.8~cm thick SP-30 type aerogel and lined with Millipore. }
\end{figure}

The horizontal scan for the aerogel with nominal refractive index $n$=1.030 is
shown in Fig.~\ref{h_scan}. Recall that in this case the aerogel tray is lined 
with Millipore. The data show a summed signal of at least 11 photoelectrons at 
the center and an increase of the signal to above 12 photoelectrons at either 
edges of the detector, close to PMTs. This is not unexpected, the same behavior
 was observed for the HMS aerogel detector~\cite{Asaturyan}. 
A slight asymmetry between left (positive $x$) and right (negative $x$) can be 
seen, more enhanced in the individual PMT sides (triangles and solid squares, 
for left and right PMTs, respectively). This asymmetry is likely caused by
an imbalance of the  quantum efficiency of the left and the right side PMTs. 
The simulated results, exhibited by the lines, do not show this left-right 
imbalance as an average quantum efficiency is assumed.

\begin{figure}
\centering
\includegraphics[width=\columnwidth]{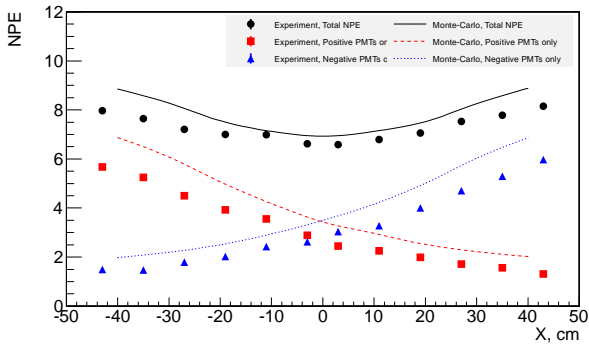}
\caption{\label{h_scan_sp20} (Color online) Dependence of signals from
  the SHMS aerogel detector on the impact point position of cosmic
  rays on the left -- right median line of the detector (see text). The tray is
  filled with 8.8~cm thick SP-20 type aerogel and lined with Millipore. }
\end{figure}

\begin{figure}[H]
\centering
\includegraphics[width=\columnwidth]{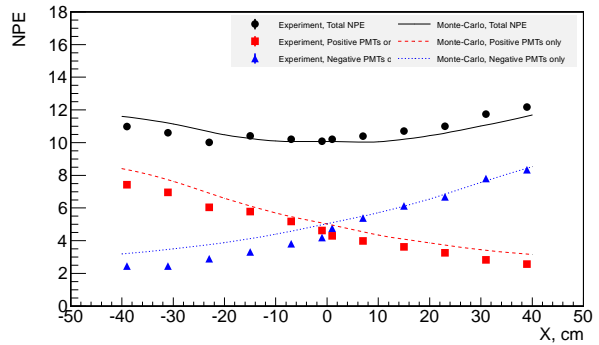}
\caption{\label{h_scan_sp15} (Color online) Dependence of signals from
  the SHMS aerogel detector on the impact point position of cosmic
  rays on the left -- right median line of the detector (see text). The tray is
  filled with 8.8~cm thick SP-15 type aerogel and lined with 1 mm Gore
  reflective material. }
\end{figure}

A similar dependence on the lateral coordinate was obtained for the aerogel 
detector when using the aerogel trays as filled with refractive index $n$=1.020
 and 1.015, SP-20 (Fig.~\ref{h_scan_sp20}) and SP-15 (Fig.~\ref{h_scan_sp15}), 
respectively.
The data for SP-20 show a signal amplitude of $\sim$7 photoelectrons in the 
center, consistent in magnitude but slightly below what one would expect based 
on the amplitude measured with SP-30 aerogel, as illustrated by the comparison 
with the predictions from simulations (solid lines). 
An increase to above 8 photoelectrons can be observed at the right and the left
 edge of the detector. For the SP-20 case, the performance overall is below the
 expectation based on the simulations.

The SP-15 data show a signal amplitude of $\sim$10 photoelectrons at the center
 of the detector. This is much higher than what one would naively expect from 
comparing to the results of the SP-30 and SP-20 tests. The SP-15 aerogel tray 
is lined with 1 mm Gore diffuse reflector material which may have increased the
 signal, but this can not account for all the increase noticed. 
The simulations shown in addition assume an aerogel absorption length of 
an order of magnitude higher than the 90 cm used for the SP-30 and SP-20 
simulations. 

Note that the SP-15 and SP-11 aerogels were manufactured by a different 
manufacturer (Japanese Fine Ceramics Center) than the older SP-20 and SP-30 
aerogels, possibly using a different production method. The larger signal 
amplitude may thus be partially due to improved quality of the aerogel material,
as observed by our measurements of high absorption lengths (low light absorption)
for the new aerogel.
Again, an increase in signal closer to the PMTs can be seen, from $\sim$ 11 
photoelectrons on the right (positive PMT) up to $\sim$12 photoelectrons on the
 left (negative PMT)

In general, if a constant threshold on the signal is used, a coordinate 
dependence of the summed signal impacts both detection efficiency and 
rejection capability in the experimental data analysis. To address this issue 
and to optimize the detector's performance the threshold level can be adjusted 
to be optimal for tracks at the center of the acceptance. Alternatively, one 
could vary the threshold depending on the coordinate, and ensure uniform 
performance across the acceptance. 

\begin{figure}[H]
\centering
\includegraphics[width=\columnwidth]{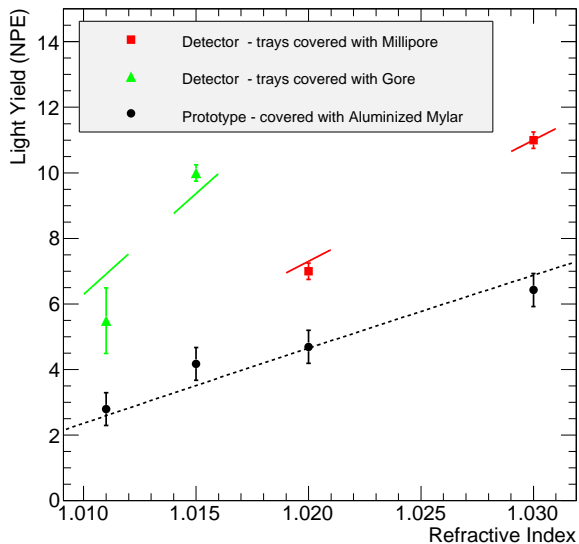}
\caption{\label{LY_detector_prototype} (Color online) Comparative study of the light yield of all four aerogel refractive indices. Aluminized mylar was used to cover the internal walls of the prototype, while Millipore covers the trays with refractive index $n$=1.030 and $n$=1.020 and Gore reflector covers the lowers index trays. The solid lines denote the expected increase in light yield measured with the full detector compared to the single counter owing predominantly to increased reflectivity of the material covering the interior walls. See text for details.}
\end{figure}

A comparative study of the light yield performance of the different aerogel refractive indices is shown in Fig.~\ref{LY_detector_prototype}. The measurements were carried out with cosmic rays using the single counter described in section~\ref{aero_study} and the full detector. For this specific test, the single counter was covered with Aluminized Mylar. The full detector trays were covered with Millipore (SP-30 and SP-20) and 1 mm thick GORE (SP-15, SP-11) reflector material. The diffusion box was covered with 60 (40)\% 3.2 (1.1) mm thick GORE reflective material. The data from the single counter measurements are well described by a fit of the form of Eq.~\ref{eq:Ncher_model2}. The light yield from the full detector is higher by about a factor of two for the higher two refractive indices. This is expected since the reflector material used for covering the interior of the full detector was optimized. Based on Fig.~\ref{fig:reflectors}, the reflectivity of aluminized mylar is about 85\% and that of Millipore about 95\% around 400nm, where the PMTs are most sensitive. Assuming ten interactions of a photon with the detector walls, this would result in light loss of 80\% with mylar compared to 40\% with Millipore reflector. The point at refractive index $n$=1.015 shows a 60\% higher light yield since the SP-15 tray was optimized with GORE reflector, which has an even higher reflectivity than Millipore. Taking into account the enhancement due to the higher reflectivity of GORE and Millipore to mylar as shown in Fig.~\ref{fig:reflectors}, the light yield data follow the expected trend as a function of refractive index. 

\section{SUMMARY}
\label{summary}

In summary, we have constructed a threshold Aerogel \v{C}erenkov detector with 
nominal refractive indices $n$=1.030, 1.020, 1.015, and 1.011, that can be 
used in the SHMS magnetic spectrometer in Hall C at JLab for experiments 
requiring kaon particle identification. 
The detector consists of one diffusion box and four exchangeable aerogel trays 
each loaded with a different nominal refractive index.

The aerogel material of the higher refractive indices ($n$=1.030 and 
$n$=1.020) was originally manufactured by Matsushita and used in the BLAST 
experiment.
The lower two refractive indices ($n$=1.015 and $n$=1.011) were obtained from 
the Japanese Fine Ceramics Center.
The optical quality of all aerogel material, such as index of refraction, 
transmittance and light yield was studied in detail. 
We found that these aerogel materials are generally of very good quality and 
there is no evidence for quality changes of the previously used material. 

The photosensors were also characterized carefully, in particular their gain 
and quantum efficiency. 
The Photonis PMTs model XP4500 has a high gain, but shows a $\sim$30--40\% lower quantum 
efficiency than the flat-faced PMTs model XP4572. For the final assembly 14 PMTs (XP4572) 
were selected from a pool 25 five-inch diameter PMTs based on the criteria of 
high quantum efficiency, low noise, high gain at relatively low high voltage, 
and good single photoelectron resolution.

The light collection performance of the detector was tested with 
cosmic rays and proton beams. The detector signal shows good uniformity along 
the vertical (Y) coordinate of the detector surface, but has a significant
dependence in the horizontal (X) direction. 
Possible optimization of this include a variable threshold and an optimized 
selection of the PMTs installed on the right and left side of the detector. 

The mean number of photo-electrons in saturation for the tray filled
with $n$=1.030 ($n$=1.020) refractive index aerogel is $\sim$12 ($\sim$8) which
is close to MC simulation.
For the trays filled with $n$=1.015 and $n$=1.011 refractive index aerogel, high number of photoelectrons were obtained with the use of higher reflectivity material to cover the tray, $\sim$10 and $\sim$5.5 respectively.
This result could be fully reproduced by our Monte Carlo simulation by also assuming
the aerogel absorption length on the order of 220 cm.

These results were obtained using Gore diffusive reflector material and 
optimizing the configuration of PMTs installed on either side of the detector. 
Using a lining of the \v{C}erenkov aerogel detector of Gore diffusive reflector
(with reflectivity above 99\%) rather than the more commonly used Millipore
(with reflectivity of 96\%) drastically improved the performance of the 
detector, by $\sim$35\%.

\vspace{0.4in}

\centerline{ACKNOWLEDGMENTS}

This work was supported in part by NSF grants PHY-1039446 and PHY-1306227, the 
CUA physics department and the Vitreous State Laboratory (VSL). In particular, 
the authors wish to thank Marek Brandys, Eric Fisher, and David Horton from 
the VSL for their expertise and support in the construction of the detector.
The detector benefited greatly from components graciously provided by both
the MIT-BLAST collaboration (aerogel materials and PMTs) and Hall C (PMTs).
We explicitly are greatful to Ricardo Alarcon and Richard Milner for help 
provided to acquire and transport the MIT-BLAST aerogel detector components.
We would also like to thank Carl Zorn from the Detector Group of the Jefferson 
Lab Physics Division, the Hall C engineering staff, in particular Bert Metzger 
for help with the design of the detector and expertise during assembly, and
Walter Kellner and the Hall C technical staff, Joe Beaufait, and Brad Sawatzky, as well as
Mariana Khachatryan from ANSL for help during various stages of the work.
The Southeastern Universities Research Association operates the Thomas 
Jefferson National Accelerator Facility under the U.S. Department of Energy
contract DEAC05-84ER40150.


\section{Appendix}

\subsection{Details on the Monte Carlo Simulation of the Detector}
\label{mc_sim}

The program for simulation of the SHMS aerogel detector is based on
the GEANT4 package \cite{geant}, version 9.2. It includes sampling of
incoming particles at the focal plane of the spectrometer and passing
them through the material of the detectors before the aerogel counter;
\v{C}erenkov light generation in the aerogel and its propagation up to
the PMT photocathodes; and signal generation from the PMTs. Interaction of
the particles with materials is modeled within the framework of the 
Quark-Gluon String Precompound physics model (QGSP), and the standard
GEANT4 optics model is used for the light generation and propagation.
 
The total thickness of the material from the SHMS detectors located between
spectrometer magneto-optical focal plane and the aerogel \v{Ce}erenkov
detector amounts to 3.7 g/cm$^2$ (see Table~\ref{shms-mat}).
The front wall of the aerogel detector adds $\sim$0.2 g/cm$^2$. 
The density of the aerogel is derived from refractive index $n$ using  
$(n-1)/0.21$ (g/cm$^3$) \cite{buzykaev}. 
For example, 9 cm of $n$=1.030 aerogel material would amount to an
effective thickness of $\sim$1.3~g/cm$^2$.

\begin{center}
\begin{table}
\caption{\label{shms-mat} The materials between the SHMS focal plane and the
aerogel \v{C}erenkov detector. The listed positions are at the fronts of 
components, as measured from the central point (particle momentum
equals the set spectrometer momentum) magneto-optical focal plane.}
{\centering  \begin{tabular}{|c|c|c|c|c|c|}
\hline
component               & material       & position  & thickness \\
                        &                &  (cm)     & $({\rm g}/{\rm cm}^2)$ \\
\hline
DC2 gas                  & Ethane/Ar     &   40       & 0.005 \\
DC2 foils                &  Mylar        &            & 0.025 \\
S1X hodoscope            & BC408 scint.  &   50       & 0.516 \\
S1Y hodoscope            & BC408 scint.  &   60       & 0.516 \\
Gas \v{C} - entr. window &     Al        &   80       & 0.270 \\
Gas \v{C} - gas          & $C_4F_8O$      &            & 0.975 \\
Gas \v{C} - mirror       &     glass      &            & 0.720 \\
Gas \v{C} - mirror support & carbon fiber &            & 0.180 \\
Gas \v{C} - exit window    & Al           &   180      & 0.270 \\
Air                        & Air at STP   &            & 0.181 \\
\hline
Total                      &              &            & 3.658 \\
\hline
\end{tabular}\par}
\end{table}
\end{center}

The aerogel material is characterized by a constant refractive index,
and by wavelength-dependent absorption and scattering lengths,
$\Lambda_A$ and $\Lambda_S$. The last 2 parameters were modeled in the simulation based
on studies of aerogel optical properties carried out
at DESY \cite{Aschenauer}. For light of wavelengths $\lambda >$300 nm the 
transmittance through the aerogel is described to good accuracy by the Hunt 
formula \cite{Hrubesh} which assumes constant absorption (constant $\Lambda_A$)
 and Rayleigh scattering ($\Lambda_S \sim \lambda^4$). 
Below 300 nm $\Lambda_A$ dramatically decreases, and $\Lambda_S$ changes less 
rapidly.

Absorption and scattering lengths of the aerogel used in the simulations 
include a constant absorption length for $\lambda >$ 300 nm linearly decreasing
 to 1 cm at $\lambda\approx$ 200 nm in accordance with Ref.~\cite{Aschenauer}. 
The constant value for $\lambda >$ 300 nm was tuned using data from cosmic 
tests with the SHMS aerogel \v{C}erenov detector 
(see section~\ref{cosmic_tests_detector})
and optimized for the optimum sensitivity of our PMTs, around 450 nm.
A value of $\Lambda_A$=90 cm gives the best agreement between simulation and 
data for SP-30 and SP-20 aerogels. This value is also consistent with our direct measurement of the absorption length at 450 nm as shown in Fig.~\ref{fig:aero_spall_absCoeff}. 
For the lower refractive index aerogels, 
which were produced by a different vendor, a larger value seems to be more 
suitable, we find $\Lambda_A \sim$220 cm.
The scale of $\Lambda_S$ is of less importance for the simulation accuracy of 
the detector's performance. For the purpose of the present simulations it was 
derived from the average value for the Hunt clarity factor $Ct$
obtained in Ref.~\cite{Aschenauer}.

Reflectivity of the Millipore paper for $\lambda \ge$ 315 nm is taken
from \cite{Benot}. Below 315 nm it is linearly extrapolated to 50\% at
$\lambda$ = 190 nm. The Millipore reflectance is taken pure (100\%) Lambertian.
The reflectance of Gore material is taken from Ref.\cite{Gore}. 

The geometry of the PMT spherical window together with adjacent
photocathode is coded using dimensions of the XP4500 and XP4572 PMT from the
vendor~\cite{photonis}. The mean quantum efficiency of the bialcali
photocathode is derived from typical spectrum of radiant sensitivity, also 
from the vendor. 



\begin{thebibliography}{99}


\bibitem{NSAC15} The 2015 Long Range Plan for Nuclear Science, \url{http://science.energy.gov/~/media/np/nsac/pdf/2015LRP/2015_LRPNS_091815.pdf} (2015)

\bibitem{mckeown15} R. D. McKeown, Jefferson Lab 12 GeV Science Program, Int. J. Mod. Phys. Conf. Ser. \textbf{37} (2015)

\bibitem{dudek12} Dudek, J \emph{et al.}, Physics opportunities with the 12 GeV upgrade at Jefferson Lab, Eur. Phys. J. A \textbf{48} 187 (2012)

\bibitem{block08} H. P. Blok \emph{et al.}, Charged pion form factor between Q$^2$=0.60 and 2.45 GeV$^2$. I. Measurements of the cross section for the $^1$H(e,e$^\prime\pi^+$)n reaction, Phys. Rev. C \textbf{78}, 045202 (2008)

\bibitem{brindza09} S.~Lassiter, P.D. Brindza, \emph{et al.}, Design of the Super Conducting Super High Momentum Spectrometer (SHMS) for the JLAB 12 GeV Upgrade,  IEEE T APPL SUPERCON \textbf{19-3} (2009)

\bibitem{E12-07-105} Scaling Study of the L-T Separated Pion Electroproduction 
Cross Section at 11 GeV, Jefferson Lab Experiment E12-07-105, 
T.~Horn and G.~M.~Huber spokespersons.

\bibitem{E12-09-011} Studies of the L-T Separated Kaon Electroproduction Cross 
Section from 5-11 GeV, Jefferson Lab Experiment E12-09-011,  
T.~Horn, G.~M.~Huber and P.~Markowitz spokespersons.

\bibitem{horn16} T.~Horn and C.~D.~Roberts, The pion: an enigma within the Standard Model, J. Phys. G: Nucl. Part. Phys. \textbf{43}, 073001 (2016)

\bibitem{favart16} L. Favart, M. Guidal, T. Horn, P. Kroll, Eur. Phys. J. A, Vol. 52, Issue 6, 158 (2016).

\bibitem{E12-09-017} Transverse Momentum Dependence of Semi-Inclusive Pion 
Production, Jefferson Lab Experiment E12-09-017,  
P.~Bosted, R.~Ent, E.~R.~Kinney, and H.~Mkrtchyan spokespersons.

\bibitem{E12-06-104} Measurement of the Ratio $R=\sigma_L/\sigma_T$ in 
Semi-Inclusive Deep-Inelastic regimes, Jefferson Lab Experiment E12-06-104, 
P.~Bosted, R.~Ent, E.~R.~Kinney and H.~Mkrtchyan spokespersons.
 
\bibitem{E12-06-107} The Search for Color Transparency at 12 GeV, Jefferson 
Lab Experiment E12-06-107,  D.~Dutta and R.~Ent spokespersons.

\bibitem{poelz86} G.~Poelz, Aerogel Cherenkov counter at DESY, 
Nucl. Instrum. Meth. A \textbf{248} (1986) 118-129.

\bibitem{carlson86} P. Carlson, Aerogel Cherenkov counters: Construction 
principles and applications, 
Nucl. Instrum. Meth. A \textbf{248} (1986) 110-117.

\bibitem{Benot} M.~Benot and J.~Carlson, Test of large Cerenkov detector with 
silica aerogel as radiator, Nucl. Instrum. Meth. A \textbf{154} (1978) 253-260

\bibitem{Asaturyan} R.~Asaturyan \emph{et al.}, The aerogel threshold Cherenkov 
detector for the High Momentum Spectrometer in Hall C at Jefferson Lab, Nucl. 
Instrum. Meth. A \textbf{546} (2005) 364-374 

\bibitem{alcorn04} J.~Alcorn \emph{et al.}, Basic instrumentation for Hall A at 
Jefferson Lab, Nucl. Instrum. Meth. A \textbf{522} (2004) 294-346.

\bibitem{millipore} Millipore Corporation, 80 Ashly Road, Bedford, MA 01730, 
http://www.millipore.com/.

\bibitem{Blast} D.~Hasell \emph{et al.}, The BLAST experiment, Nucl. Instrum. 
Meth. A \textbf{603} (2009) 247-262;
B.~Tonguc \emph{et al.}, The BLAST \u{C}erenkov Detector, Nucl. Instrum. 
Meth. A \textbf{553} (2005) 364-9. 

\bibitem{adachi} I.~Adachi \emph{et al.}, Study of highly transparent silica 
aerogel as a RICH radiator, 
Nucl. Instrum. Meth. A \textbf{553} (2005) 146-151.

\bibitem{Aschenauer} E.~Aschenauer \emph{et al.}, Optical characterization of 
$n$=1.03 silica aerogel used as radiator in the RICH of HERMES, 
Nucl. Instrum. Meth. A \textbf{440} (2000) 338-347.

\bibitem{Matsushita} Matsushita Electric Works Ltd, 1048 Kadoma, Kadoma-shi,
Osaka 571, Japan.

\bibitem{Jackson} J.~D.~Jackson, Classical Electrodynamics, 3rd edition,
John Wiley and Sons, New York, 1998.

\bibitem{doug98} D.~Higinbotham, Diffusely reflective aerogel Cherenkov 
detector simulation techniques, Nucl. Instrum. Meth. A 414 (1998) 332-339.

\bibitem{buzykaev} A.~R.~Buzykaev \emph{et al.}, Measurement of optical 
parameters of aerogel, Nucl. Instrum. Meth. A \textbf{433} (1999) 396-400.

\bibitem{NIST_air_refIndex} Jack A. Stone and Jay H. Zimmerman, Wavelength in 
Ambient Air and Refractive Index of Air Based on Ciddor Equation,\\
 \url{http://emtoolbox.nist.gov/Wavelength/Ciddor.asp} (consulted on March 2014)

\bibitem{Danilyuk} A.~F.~Danilyuk \emph{et al.}, Recent results on aerogel 
development for use in Cherenkov counters, Nucl. Instrum. Meth.
A \textbf{494} (2002) 491-494. 

\bibitem{Barnyakov} A.~Yu.~Barnyakov \emph{et al.}, Aerogel Cherenkov Counter 
for the SND Detector, SND Symposium, Stanford, California, 3-6 April 2006. 

\bibitem{Allada} K.~Allada, Ch.~Hurlbut, L.~Ou, B.~Schmookler, A. Shahinyan, 
B.~Wojtsekhowski, PMT signal increase using a wavelength shifting paint, Nucl. Instrum. Meth. A \textbf{782} 87-91 (2015).

\bibitem{Gogami} T.~Gogami \emph{et al.}, Bucking coil implementation on PMT 
for active canceling of magnetic field, Nucl. Instrum. Meth. A \textbf{729} (2013) 816-824,
arXiv:1307.0896v1, [physics.ins-det], 3 Jul 2013.

\bibitem{geant} \url{http://geant4.cern.ch/support/proc-mod-catalog/physics-lists}.

\bibitem{Hrubesh} L.~W.~Hrubesh, Optical Characterization of Silica 
AerogelGlass,U.C.R.L. Report no. 53794, 1987.

\bibitem{photonis} Photomultipliers, Philips, PC04.\\ 
\url{http://www.photonis.com/upload/industryscience/pdf}.

\bibitem{Nishida} S. Nishida \emph{et al.}, Nucl. Instrum. Meth. 
A \textbf{595} (2008) 150-153.

\bibitem{Coman} The Hall A Aerogel \v{C}erenkov Detector, Marius Coman, 
Master Thesis, FIU, 2000.

\bibitem{Grove} J.~E.~Grove and R.~A.~ Mewaldt, Nucl. Instrum. Meth. 
A \textbf{314} (1992) 495-503.

\bibitem{Bellunato} T.~Bellunato \emph{et al.}, Nucl. Instrum. Meth. 
A \textbf{493} (2004) 493.

\bibitem{Sumiyoshi} T.~Sumiyoshi \emph{et al.}, Nucl. Instrum. Meth.
A \textbf{433} (1999) 385-391. 

\bibitem{Gore} W.~L.~Gore \& Assosiates INC., \url{http://www.gore.com}.

\bibitem{Mkrtchyan} H.~Mkrtchyan \emph{et al.}, The lead-glass electromagnetic 
calorimeters for the magnetic spectrometers in Hall C at Jefferson Lab., 
Nucl. Instrum. Meth. A \textbf{719} 2013 85-100.

\bibitem{Grzonka} D.~Grzonka \emph{et al.}, Search for polarization effects in 
the antiproton production process, 
Acta Physics Polonica Series B, 01/2015 46(1); 
arXiv:1501.05730 [physics.ins-det] 23 Jan 2015, 12 pages.

\bibitem{G0} D.~Androi\v{c}  \emph{et al.}, Nucl. Instrum. Meth.
A \textbf{646} (2011) 59-66. 

\bibitem{ElJen} ElJen Technology, 1300 W Broodway, Sweetwater, TX 79556, USA;\\
\url{www.eljentechnology.com}

\bibitem{Seltzer-Berger} M.~J.~Berger and S.~M.~Seltzer, NASA Technical 
Report SP-3012, Washington D.~C., 1964.

\bibitem{Evenson} P.~Evenson, 14th International Cosmic Ray Conference, 
Conference Papers, vol. 9, p.3177, Munich, 1975.


\end{thebibliography}
\end{document}